\documentclass[11pt]{article}
\usepackage{amssymb,macros2e,graphicx}
\usepackage{times} 
\skip\footins 39pt plus 4pt minus 2pt
\def\footnoterule{\kern-19pt\hrule width.5in\kern18.6pt}%
\newcommand{\dotsb}{\ldots}

\newcommand{\maxmin}{\mbox{$\mbox{max}\atop \mbox{min}$}}
\newcommand{\tr}{\mbox{tr}}
\newcommand{\ind}[1]{{\mathrm{#1}}}
\newcommand{\be}{\begin{equation}}
\newcommand{\ee}{\end{equation}}
\newcommand{\bea}{\begin{eqnarray}}
\newcommand{\eea}{\end{eqnarray}}

\newcommand{\half}{\frac{1}{2}}
\newcommand{\ts}{\hskip0.1ex\raisebox{-1ex}[0ex][0.8ex]{\rule{0.1ex}{2.75ex}\hskip0.2ex}}
\newcommand{\lts}{
   {\raisebox{0ex}{\parbox{0.5ex}{
      \setlength{\unitlength}{0.5ex}
      \begin{picture}(1,12)
         \thinlines
         \put(0.5,-0.25){\line(0,1){12}}
      \end{picture}
   }}}
}
\newcommand{\fz}{\mathfrak{z}}
\begin{document}
%
\newif\ifpdf
	\ifx\pdfoutput\undefined
	\pdffalse 
	\newcommand{\fig}[2]{\includegraphics[width=#1]{./figures/#2.eps}}
	\newcommand{\Fig}[1]{\includegraphics[width=\columnwidth]{./figures/#1.eps}}
         \newlength{\bilderlength} 
	\newcommand{\bilderscale}{0.22}
	\newcommand{\storebilderscale}{\bilderscale}
	\newcommand{\bilderskip}{\hspace*{0.8ex}}
	\newcommand{\textdiagram}[1]{%
	\renewcommand{\bilderscale}{0.16}%
	\diagram{#1}\renewcommand{\bilderscale}{\storebilderscale}}
	\newcommand{\vardiagram}[2]{%
	\renewcommand{\bilderscale}{#1}%
	\diagram{#2}\renewcommand{\bilderscale}{\storebilderscale}}
	\newcommand{\diagram}[1]{%
	\settowidth{\bilderlength}{\bilderskip%
	\includegraphics[scale=\bilderscale]{./figures/#1.eps}\bilderskip}%
	\parbox{\bilderlength}{\bilderskip%
	\includegraphics[scale=\bilderscale]{./figures/#1.eps}\bilderskip}}
	\newcommand{\Diagram}[1]{%
	\settowidth{\bilderlength}{%
	\includegraphics[scale=\bilderscale]{./figures/#1.eps}}%
	\parbox{\bilderlength}{%
	\includegraphics[scale=\bilderscale]{./figures/#1.eps}}}
\else
	\pdfoutput=1 
	\newcommand{\fig}[2]{\includegraphics[width=#1]{./figures/#2.pdf}}
	\newcommand{\Fig}[1]{\includegraphics[width=\columnwidth]{./figures/#1.pdf}}
	\newlength{\bilderlength} 
	\newcommand{\bilderscale}{0.22}
	\newcommand{\storebilderscale}{\bilderscale}
	\newcommand{\bilderskip}{\hspace*{0.8ex}}
	\newcommand{\textdiagram}[1]{%
	\renewcommand{\bilderscale}{0.16}%
	\diagram{#1}\renewcommand{\bilderscale}{\storebilderscale}}
	\newcommand{\vardiagram}[2]{%
	\renewcommand{\bilderscale}{#1}%
	\diagram{#2}\renewcommand{\bilderscale}{\storebilderscale}}
	\newcommand{\diagram}[1]{%
	\settowidth{\bilderlength}{\bilderskip%
	\includegraphics[scale=\bilderscale]{./figures/#1.pdf}\bilderskip}%
	\parbox{\bilderlength}{\bilderskip%
	\includegraphics[scale=\bilderscale]{./figures/#1.pdf}\bilderskip}}
	\newcommand{\Diagram}[1]{%
	\settowidth{\bilderlength}{%
	\includegraphics[scale=\bilderscale]{./figures/#1.pdf}}%
	\parbox{\bilderlength}{%
	\includegraphics[scale=\bilderscale]{./figures/#1.pdf}}}
	\pdftrue
\fi

%

%
\newcommand{\sgn}{{\mathrm{sgn}}}
\newcommand{\rme}{{\mathrm{e}}}
\newcommand{\rmd}{{\mathrm{d}}} 
\newcommand{\nn}{\nonumber}\newcommand {\eq}[1]{(\ref{#1})}
\newcommand {\Eq}[1]{Eq.\hspace{0.55ex}(\ref{#1})}
\newcommand {\Eqs}[1]{Eqs.\hspace{0.55ex}(\ref{#1})}
\newcommand{\E}{\epsilon}

\def\true{true}
\newsavebox{\bilderbox}
\newlength{\bilderhelp}
\newsavebox{\bilderone}
\newlength{\bilderonelength}
\newsavebox{\bildertwo}
\newlength{\bildertwolength}
%
\newcommand{\bild}[1]{\fboxsep0mm%
\sbox{\bilderbox}{
{\includegraphics[scale=\bilderscale]{#1}}}%
\settowidth{\bilderlength}{\usebox{\bilderbox}}%
\parbox{\bilderlength}{\usebox{\bilderbox}}}
\newcommand{\savebild}[3]{\newsavebox{#2}%
\sbox{#2}{\bild{#3}}\newcommand{#1}{%
\ensuremath{\,\mathchoice{\usebox{#2}}%
{\settowidth{\bilderhelp}{\scalebox{0.7}{\usebox{#2}}}%
\parbox{\bilderhelp}{\scalebox{0.7}{\usebox{#2}}}}%
{\settowidth{\bilderhelp}{\scalebox{0.5}{\usebox{#2}}}%
\parbox{\bilderhelp}{\scalebox{0.5}{\usebox{#2}}}}%
{\settowidth{\bilderhelp}{\scalebox{0.35}{\usebox{#2}}}%
\parbox{\bilderhelp}{\scalebox{0.35}{\usebox{#2}}}}%
\,}}}
\newcommand{\bilderdiagram}[4]{{%
\mathchoice{
\sbox{\bilderone}{\ensuremath{\displaystyle#1}}%
\sbox{\bildertwo}{\ensuremath{\displaystyle#2}}%
\settoheight{\bilderonelength}{\ensuremath{\usebox{\bilderone}}}%
\settoheight{\bildertwolength}{\ensuremath{\usebox{\bildertwo}}}%
\left#3\!\usebox{\bilderone}{\rule{0mm}{\bildertwolength}}\right.%
\hspace*{-0.5ex}\!\left|\!{\rule{0mm}{\bilderonelength}}%
\usebox{\bildertwo}\!\right#4}%
{
\sbox{\bilderone}{\ensuremath{\textstyle#1}}%
\sbox{\bildertwo}{\ensuremath{\textstyle#2}}%
\settoheight{\bilderonelength}{\ensuremath{\usebox{\bilderone}}}%
\settoheight{\bildertwolength}{\ensuremath{\usebox{\bildertwo}}}%
\left#3\!\usebox{\bilderone}{\rule{0mm}{\bildertwolength}}\right.%
\hspace*{-0.35ex}\!\left|\!{\rule{0mm}{\bilderonelength}}%
\usebox{\bildertwo}\!\right#4}%
{
\sbox{\bilderone}{\ensuremath{\scriptstyle#1}}%
\sbox{\bildertwo}{\ensuremath{\scriptstyle#2}}%
\settoheight{\bilderonelength}{\ensuremath{\usebox{\bilderone}}}%
\settoheight{\bildertwolength}{\ensuremath{\usebox{\bildertwo}}}%
\left#3\!\usebox{\bilderone}{\rule{0mm}{\bildertwolength}}\right.%
\hspace*{-0.1ex}\!\left|\!{\rule{0mm}{\bilderonelength}}%
\usebox{\bildertwo}\!\right#4}%
{
\sbox{\bilderone}{\ensuremath{\scriptscriptstyle#1}}%
\sbox{\bildertwo}{\ensuremath{\scriptscriptstyle#2}}%
\settoheight{\bilderonelength}{\ensuremath{\usebox{\bilderone}}}%
\settoheight{\bildertwolength}{\ensuremath{\usebox{\bildertwo}}}%
\left#3\!\usebox{\bilderone}{\rule{0mm}{\bildertwolength}}\right.%
\hspace*{-0.1ex}\!\left|\!{\rule{0mm}{\bilderonelength}}%
\usebox{\bildertwo}\!\right#4}%
}}
\newcommand{\MOPE}[2]{\bilderdiagram{#1}{#2}{(}{)}}
\newcommand{\DIAG}[2]{\bilderdiagram{#1}{#2}{<}{>}}
\newcommand{\DIAGindhelp}[4]{%
\sbox{\bilderbox}{\ensuremath{#4\bilderdiagram{#1}{#2}{<}{>}}}%
\settowidth{\bilderlength}{\rotatebox{90}{\ensuremath{\usebox{\bilderbox}}}}%
\ensuremath{\usebox{\bilderbox}_{\hspace*{-0.162\bilderlength}#3}}}
\newcommand{\DIAGind}[3]{%
\mathchoice{\DIAGindhelp{#1}{#2}{#3}{\displaystyle}}%
{\DIAGindhelp{#1}{#2}{#3}{\textstyle}}%
{\DIAGindhelp{#1}{#2}{#3}{\scriptstyle}}%
{\DIAGindhelp{#1}{#2}{#3}{\scriptscriptstyle}}}
\newcommand{\reducedbildheightrule}[2]{{%
\mathchoice{\settowidth{\bilderlength}{\rotatebox{90}{\ensuremath{\displaystyle#1}}}%
\parbox{0mm}{\rule{0mm}{#2\bilderlength}}}%
{\settowidth{\bilderlength}{\rotatebox{90}{\ensuremath{\textstyle#1}}}%
\parbox{0mm}{\rule{0mm}{#2\bilderlength}}}%
{\settowidth{\bilderlength}{\rotatebox{90}{\ensuremath{\scriptstyle#1}}}%
\parbox{0mm}{\rule{0mm}{#2\bilderlength}}}%
{\settowidth{\bilderlength}{\rotatebox{90}{\ensuremath{\scriptscriptstyle#1}}}%
\parbox{0mm}{\rule{0mm}{#2\bilderlength}}}}}
\newcommand{\bildheightrule}[1]{\reducedbildheightrule{#1}{1}}
\newcommand{\reducedbild}[2]{%
{\settowidth{\bilderhelp}{#1}%
\setlength{\bilderhelp}{#2\bilderhelp}%
\parbox{\bilderhelp}{\scalebox{#2}{#1}}}}
%
%


\centerline{\sffamily\bfseries\large  The 4-loop $\beta$-function
in the 2D Non-Abelian Thirring model,}
\centerline{\sffamily\bfseries\large  
 and comparison with its conjectured ``exact''  form}

\bigskip  
\centerline{\sffamily\bfseries\normalsize Andreas W. W. Ludwig{$^{1}$}
 and Kay J\"org Wiese{$^2$}}
\bigskip 
\centerline{{$^1$}Physics Department, University of California at
Santa Barbara, Santa Barbara, CA 93106, USA}
\centerline{$^{2}$KITP, Kohn Hall, University of
California at 
Santa Barbara, Santa Barbara, CA 93106, USA}
\medskip 

\medskip 
\centerline{\small Nov.~23, 2002}

\noindent \rule{\textwidth}{0.3mm} \smallskip \leftline{\bfseries
Abstract} Recently, B.~Gerganov, A.~LeClair and M.~Moriconi
[Phys. Rev. Lett. {\bf 86} (2001) 4753] have proposed an ``exact''
(all-orders) 
$\beta$-function for 2-dimensional conformal field theories with
Kac-Moody current-algebra symmetry at any level $k$, based on a Lie
group ${\cal G}$, which are perturbed by a current-current
inter\-action.  This theory is also known as the Non-Abelian Thirring
model.  We check this conjecture with an explicit calculation of the
$\beta$-function to 4-loop order, for the classical groups ${\cal G}=
\mbox{SU}(N)$, $\mbox{SO}(N)$ and $\mbox{SP}(N)$ at level $k=0$.  We
find a contribution at 4-loop order, proportional to a higher-order
group-theoretical invariant, which is incompatible with the proposed
$\beta$-function in {\em all} possible regularization schemes.

\noindent \rule{\textwidth}{0.3mm}


\small
\tableofcontents

\vfill
\eject
\normalsize
\setcounter{page}{2}

\section{Introduction}\label{intro} 

Perturbations of conformal field theories (CFT) in two dimensions (2D)
have been a very active topic of study for a long time.  The focus of
this paper are 2D conformal field theories possessing Kac-Moody
current algebra (or: ``Affine Lie algebra'') symmetry
\cite{KnizhnikZamolodchikov1984} associated with a Lie group ${\cal
G}$, which are perturbed by a bilinear in the Noether-current (i.e.\
by a ``left-right current-current bilinear interaction'').
Non-abelian Thirring models
\cite{DashenFrishman1973,DashenFrishman1975} and Gross-Neveu models
\cite{GrossNeveu1974} are much studied examples.  Typically, such
perturbations are (marginally) relevant and generate a mass scale; in
these cases the long-distance (infrared) behavior of these theories is
that of a massive field theory. However, generalizations of these
theories may exhibit\cite{LudwigFisherShankarGrinstein1994} massless
long-distance fixed points in a certain zero species (``replica'')
limit, or, which can be seen to be equivalent\cite{EfetovBook}, when
the symmetry group ${\cal G}$ is replaced by a certain supergroup.
Such theories are of great interest in condensed matter physics, since
the mentioned ``zero-species'', or equivalently, supersymmetry
generalizations describe (de-)localization transitions known to occur
in dirty, i.e.\ disordered, non-interacting electronic systems in two
spatial dimensions, subject to static random impurities.  Indeed, the
aim of \cite{LudwigFisherShankarGrinstein1994} was to study the
integer quantum Hall plateau transition\footnote{in the absence of
long-range electron-electron interactions}, and to provide an
alternative to the formulation given by Levine, Libby and
Pruisken\cite{LevineLibbyPruisken1983,LevineLibbyPruisken1984a,LevineLibbyPruisken1984b,LevineLibbyPruisken1984c,Khmelnitskii1983}
in terms of a strongly coupled non-linear sigma model with a
topological term.  More recently it was
recognized\cite{Zirnbauer1996,AltlandZirnbauer1997,SenthilFisherBalentsNayak1998,SenthilFisher2000}
that disordered superconductors (and other systems) provide an entire
new arena capable of exhibiting (de-)localization transitions of a
similar kind (albeit in entirely new universality classes).  Since
then, the study of (de-)localization transitions in non-interacting
quantum systems in 2D has seen an immense surge of research activity.

A general understanding of the strong-coupling (long-distance)
behavior of 2D Kac-Moody current algebras perturbed by current-current
interactions would be a valuable tool to describe a number of such
transitions.  Indeed, a good understanding of such perturbations has
been achieved in a few cases
\cite{LudwigFisherShankarGrinstein1994,MudryChamonWen1996,ChamonMudryWen1996,GuruswamyLeClairLudwig2000}.
However, in general, this is not the case, to date.

An intriguing conjecture has recently been advanced by Gerganov,
LeClair and Moriconi\cite{GerganovLeClairMoriconi2001}, who consider,
as above, a general Kac-Moody current algebra conformal field theory
with symmetry group ${\cal G}$ at any level $k$, perturbed by
right-left current bilinears. (The perturbations they discuss may also
be anisotropic, or involve a supergroup, but this will not be
important for the arguments presented in this article, which focuses
entirely on the isotropic situation and bosonic groups.)  Their paper
builds on earlier work by Kutasov\cite{Kutasov1989}, who computed the
renormalization group (RG) $\beta$-function for the isotropic
case\footnote{This theory is renormalizable with a single coupling
constant.} of a (bosonic) symmetry group ${\cal G}$ to leading order
in the large level $k$ of the current algebra.  The authors of
\cite{GerganovLeClairMoriconi2001} argued that Kutasov's result be
exact for {\it any} value of the level $k$, in a particular
regularization scheme.  Specifically, for the isotropic case, the
authors of Ref.~\cite{GerganovLeClairMoriconi2001} conjecture that the
{\it exact} $\beta$-function for the coupling $g$ be given by
\footnote{ $l:=\ln (a/L)$ is the RG-flow parameter, and $a$ and
$L$ are the UV and IR cutoffs, respectively.}
\begin{equation}
\label{conjecturedbetafct}
 \beta (g):= {\rmd g \over \rmd l} = {1\over 2} {
C_{2} g^2 \over (1 + k g /4)^2 }\ ,
\end{equation}
where $ C_{2}$ is the eigenvalue of the quadratic Casimir operator in
the adjoint representation of the symmetry group ${\cal G}$.  Clearly,
the notion of an {\em exact} $\beta$-function is delicate due to its
dependence on the regularization scheme.  (The contributions to the
$\beta$-function beyond 2-loop order are scheme dependent.)
Ref.~\cite{GerganovLeClairMoriconi2001} appears to be working in some
scheme related to the left-right factorization of the underlying CFT.
However, an explicit cut-off procedure, within which
Eq.~(\ref{conjecturedbetafct}) is to be valid, is not specified in
more detail in \cite{GerganovLeClairMoriconi2001}. The authors
indicate that certain checks to 3-loop order were performed.  Checks
beyond three loop order have never been performed, to our knowledge.
The 3-loop $\beta$-function within dimensional regularization has also
been discussed in Refs.~\cite{BennettGracey1999,AliGracey2001}.

For the case where a symmetry ${\cal G}=\mbox{SU}(2)$ is broken down
to $\mbox{U}(1)$ by a purely imaginary easy-axis anisotropy, and for
level $k=1$, the above conjecture (more precisely, an appropriate
generalization) reproduces\cite{BernardLeClair2001} known exact
results\cite{Nienhuis1984,FendleySaleurZamolodchikov1993a,FendleySaleurZamolodchikov1993b}.
The massless RG flow of the resulting non-unitary theory interpolates
between ultraviolet (UV) and massless infrared (IR) fixed
points. (Both lie on the line of free scalar field theories with
central charge $c=1$ but with different compactification
radii\footnote{Due to the non-unitarity of the theory, this is not in
conflict with Zamolodchikov's c-Theorem\cite{Zamolodchikov1986}.}).
In this case, the conjectured $\beta$-function reproduces correctly
the exactly known universal relationship between the exact scaling
exponents (i.e.\ the slopes of the $\beta$-function) at the IR and the
UV fixed points.  This is the only universal information contained in
any $\beta$-function describing this flow.

On the other hand, the conjectured form of the $\beta$-function,
when appropriately generalized to supergroups, and to the anisotropic
case, has recently been applied 
\cite{BernardLeclair2002} to theories
describing disordered systems, of the kind mentioned above.  Here
however, certain problems were encountered: Integration of the
conjectured RG equations led to flows which reached a singularity
after a finite scale transformation, which appears to be an
unacceptable result.

Motivated by these inconclusive results concerning the validity of the
conjecture, we were led to check the conjec\-tured form of the $\beta
$-function by explicit computation to high loop order.  We consider
the classical symmetry groups ${\cal G}= \mbox{SU}(N),\ \mbox{SO}(N)$,
and $\mbox{SP}(N)$, and the special case of {\it isotropic}
current-current interactions at level $k=0$.  Our results are
summarized in the following section. 
For all the classical groups, we find a contribution to the
$\beta$-function at 4-loop order which is incompatible with the
conjecture in {\it all} possible regularization schemes.  The
discrepancy is caused by an extra logarithmic divergence in
perturbation theory, proportional to an additional group theoretical
invariant (besides the quadratic Casimir), which first appears at
4-loop order. This divergence is not accounted for by the conjectured
form of the $\beta$-function.  Implications of our results, obtained
for level $k=0$, for the $k$-dependence of the $\beta$-function in any
scheme are discussed in the conclusion, section \ref{Conclusion}.  In
this section, we also come back to, and comment on the special case of
the anisotropic $\mbox{SU}(2)$ model mentioned above.  The reader who
wishes to skip the technical details of our calculation, which are
presented in sections \ref{Model and Method} and \ref{secCalculation},
will find a self-contained exposition of our results in sections
\ref{present results} and \ref{Conclusion}.

\section{Presentation of results}
\label{present results}
\begin{figure}[t]
\Fig{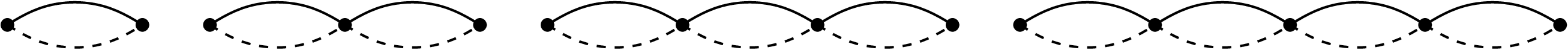}
\caption{Chain-diagrams up to 4-loop order. The number of bubbles is
the number of loops.}
\label{f:chains}
\end{figure}

\begin{figure}[tbp!]
\centerline{\fig{0.3\textwidth}{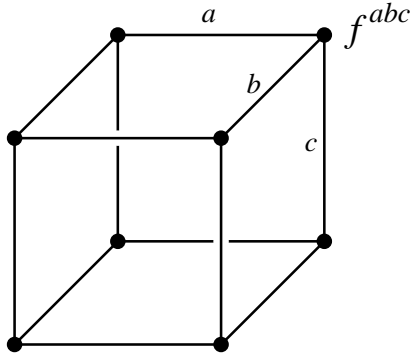}}
\caption{The group-theoretical invariant $d_{2}$ and its graphical
representation.}
\label{cube}
\end{figure}

In order to check the conjecture, we consider, as mentioned above,
 the   specific case of an isotropic perturbation
with symmetry group ${\cal G}$, and
 level  $k=0$. The conjectured $\beta$-function
(\ref{conjecturedbetafct}) then becomes
\begin{equation}\label{conj}
\beta (g)  = \frac{1}{2} C_{2} g^{2}\ .  
\end{equation}
Here $C_2$ denotes the eigenvalue of the quadratic Casimir invariant
in the adjoint representation of ${\cal G}$.  For the
classical groups, ${\cal G} = \mbox{SU}(N),\ \mbox{SO}(N)$, and
$\mbox{SP}(N)$, these  are listed, in our normalizations, in figure
\ref{grouptable}.
\begin{figure}
$$
\begin{array}{rll}
\mbox{SU} (N)\,: &\qquad  C_{2}= N\qquad  & d_{2} = \frac{3}{2}N^{2} \\
\mbox{SO} (N)\,: &\qquad  C_{2}= N-2 \qquad  & d_{2} = 24-\frac{45}{2}N+6N^{2}-\frac{3}{8}N^{3}\rule{0mm}{5mm}\\
\mbox{SP} (N)\,: &\qquad  C_{2}= \frac{N+2}{2}\qquad  & d_{2} =
\frac{3}{2} +\frac{45}{32}N+\frac{3}{8}N^{2}+\frac{3}{128}N^{3}\rule{0mm}{5mm}
\end{array}
$$
\caption{Group theoretical numbers for the classical groups, used
in the main text.}
\label{grouptable}
\end{figure}%
In sections \ref{1loop} to \ref{beta4loop} we present an
explicit perturbative calculation of the $\beta$-function up to 4-loop
order. This calculation proceeds in three steps:
\begin{enumerate}
\item [(i)] use the current-algebra to calculate the diagrams,
\item [(ii)] simplify the diagrams using elementary algebra,
\item [(iii)] evaluate the integrals, which represent the
(``Feynman'') diagrams.
\end{enumerate}
After step (i), we encounter a great number of (rather complicated
looking) diagrams. However after step (ii), we are left with only two
classes of diagrams,
\begin{enumerate}
\item  chain  diagrams  (``bubble'' diagrams),
\item  non-chain diagrams. 
\end{enumerate}
{\it Chain-diagrams} appear at loop-order 1, 2, 3, and 4. They have
the form depicted in figure \ref{f:chains}.
Each bubble comes with a factor of $g$ (the coupling-constant), with a
group-theoretical factor of $C_{2}$, and the whole chain with an
($N$-independent) integral ${\cal I}_{n}$, at $n$-loop order ($n$
bubbles in the chain).  The integral ${\cal I}_{n}$ depends on the
cutoffs $L$ (infrared) and $a$ (ultraviolet), and is a
polynomial\footnote{See (\ref{lf31}) for concreteness.} of
degree $n$ in $[\ln(\frac{L}{a} )]$, plus terms which are finite for
$L/a\to \infty$. (The fact that an $n$-loop integral is bounded by $c
\left[ \ln (\frac{L}{a} )\right]^{n}$ with some constant $c$ is a
necessity to ensure renormalizability.)  Only the leading term in
${\cal I}_n$, with the highest power of $\ln (\frac{L}{a} )$, is
universal.  This applies e.g.\ to the diverging part of the 1-loop
integral ${\cal I}_{1}$.  Thus the contribution of the chain diagrams
to the renormalization of the coupling $g$ is, at $n$-loop order (up
to a combinatorial factor)
\begin{equation}\label{n chains}
g ( C_{2}g)^{n}{\cal I}_{n}\ .
\end{equation}
{\it Non-chain diagrams} first appear at 4-loop order. They are
proportional to $d_{2}$, which is an additional group-theoretical
invariant (in the adjoint representation), independent of the
quadratic Casimir $C_{2}$. Its value for the classical groups is given
on figure \ref{grouptable}. This invariant can be constructed by
drawing a cube, where one puts a factor of $f^{ab}_{c}$ on each
corner, with one of its three indices on each adjoining edge, see
figure \ref{cube}. Finally, indices on the same edge are contracted.

Each non-chain diagram has a global divergence proportional to $[\ln
(\frac{L}{a})]$ (``single log''), and subdivergences (higher powers of
$[\ln (\frac{L}{a})]$).  However, it turns out that to the order
considered here, one can always group non-chain diagrams together into
classes, such that each class has only a global divergence, but {\em no
subdivergence}.  This means that the integral (over positions) is
proportional to $\ln (\frac{L}{a} )+ \mbox{finite}$, and that the
prefactor in front of $[\ln (\frac{L}{a})]$ is again {\em universal}.

Let us recall here that a diagram proportional to $[\ln
(\frac{L}{a})]$, i.e.\ a ``single log'', gives a finite contribution
to the $\beta$-function.  Using the above information, we thus find
the following $\beta$-function at 4-loop order
\begin{equation}
\label{beta-4-loop-unspec}
\beta (g) =   g\left[\frac{1}{2} C_{2} g + a_{2} (C_{2} g)^{2}+ a_{3}
(C_{2} g)^{3}+a_{4} (C_{2} g)^{4}   - d_{2} \frac{\pi }{240}
(6+\pi ^{2} )  g^{4} \right] + O(g^6)\ .
\end{equation}
Here, the numbers $a_{3}$ and $a_{4}$ depend on the regularization
scheme (but not on $N$ or the value of the cutoffs). In contrast,
the remaining three terms are universal.  The first
term comes from the ``single log'' of the 1-loop chain (giving the
contribution $\half C_2 g$) and the last term from that of the
non-chain diagrams (giving $- d_{2} \frac{\pi }{240} (6+\pi ^{2} )
g^{4}$).  Furthermore, we have $a_{2}=0$ since we consider level
$k=0$; this is a consequence of the universality of the
$\beta$-function up to 2 loops, and will be checked for a specific
scheme in appendix \ref{app:2loopfac}. Note that to arrive at the
result in (\ref{beta-4-loop-unspec}), {\em no specific form of the
cutoff procedure has to be chosen}.

We are now in a position to answer the question, of whether the
conjecture is compatible with our explicit calculation, in some given
cut-off scheme. To see this, consider for example the group ${\cal G}=
\mbox{SU} (N)$, where we have $C_{2}=N$, and $d_{2}={3\over 2} N^2$
(see Fig.~\ref{grouptable}).  Thus at 4-loop order, we have the following
contributions
\begin{equation}\label{4l:}
g \left[ a_{4} N^{4} g^{4} - \frac{3N^{2}}{2}\frac{\pi }{240} (6+\pi ^{2} )
g^{4}  \right] \ .
\end{equation}
Since $C_2$ and $d_2$ contain all the dependence on $N$, and since
$a_4$ is independent of $N$, there is no possible choice of $a_4$, and
thus no cutoff procedure, which cancels this term {\em for all} $N$.
This proves, that the conjecture is incorrect for all possible cutoff
procedures, for level $k=0$.  The same conclusion is arrived at for
the other choices of groups.

Let us finally give the result in a specific scheme, namely in the
scheme in which an $n$-loop chain diagram is proportional to $[\ln
(\frac{L}{a})]^{n}$, with no subleading term in $\ln (\frac{L}{a})$:
\begin{equation}\label{beta-4-loop-results}
\beta (g) = \frac{1}{2}  C_{2} g^{2} - d_{2} \frac{\pi }{240}
(6+\pi ^{2} )  g^{5} + O(g^6)\ .
\end{equation}
Indeed, this is the scheme used in the large-$N$ expansion of the
Gross-Neveu model, where the $\beta$-function becomes
quadratic\cite{Zinn}, to leading order in $1/N$.  Note that this is
compatible with our result (\ref{beta-4-loop-results}), since (upon
rescaling $g$ by $N$), the first (1-loop) term is order $O(1)$,
whereas the second (4-loop) term is order $O(1/N^2)$.

In conclusion, we have found that the conjecture is violated at 4-loop
order, and at order $1/N^{2}$ for $\mbox{SU} (N)$. For $\mbox{SO}
(N)$ and $\mbox{SP} (N)$ corrections appear at order $1/N$, as can be
seen from the table in figure \ref{grouptable}.

\medskip Let us now outline the organization of the article; the
reader wishing to skip the technical details of our paper can proceed
directly to section \ref{Conclusion}: In section \ref{Model and
Method} we introduce the model, the current-algebra and basic
notations.  Our calculations are presented in section 4: We show in
section \ref{1loop} how the Kac-Moody current-algebra is used to
successively eliminate interaction vertices from expectation values,
and how this can be used to evaluate OPE-coefficients. This is a
non-trivial task. Indeed, the raw result of this reduction procedure
depends on the order of the successive reductions and is highly
asymmetric, whereas the OPE-coefficient should be symmetric. To obtain
a more symmetric result, the raw result can be simplified by using
algebraic relations which we have baptized ``magic rules'', for their
efficiency.  This will explicitly be demonstrated in section
\ref{2loop} on the example of the 2-loop diagrams. In section
\ref{3loop} we proceed to 3-loop order, and show again how the initial
highly asymmetrically looking OPE-coefficient is simplified. As in
2-loop order, all resulting diagrams are chain-diagrams which in a
suitable scheme factorize, and thus do not give a new contribution to
the $\beta$-function. Proceeding to 4-loop order in section \ref{4loop
direct}, one finds diagrams which the magic rules are no longer able
to simplify to chain-diagrams. Due to the sheer number of initial
diagrams, namely 576, this approach is not very illuminating. In
section \ref{axialOPEs} we therefore pursue a different route: We
first calculate OPE-coefficients for ``adjoint'' perturbations $\Phi
^{a}=f^{abc}J^{b}\bar J^{c}$. We then show in section \ref{4loop
simplified} how an $n$-loop OPE-coefficient can be expressed as a
simple algebraic function times an OPE-coefficient at order $n-1$,
involving two adjoint perturbations. This allows us to identify at
4-loop order a combination of eight diagrams, which can not be
factorized as chains (non-chain diagrams). Their contribution to the
4-loop $\beta$-function is calculated analytically in appendix
\ref{4loopintegral}.  All these ingredients are collected in section
4.7, where we obtain the 4-loop $\beta$-function.  We have relegated
some basic group-theoretical relations to appendices \ref{a:algebra},
\ref{SUn} and \ref{otheralgebras}.  Conclusions and further
perspectives are offered in section \ref{Conclusion}.



\section{Model and method}
 \label{Model and Method} 
We study the
Non-Abelian Thirring Model (NATM) in two dimensions.  The model may be
defined as a perturbation of a 2D conformal field theory, with action
$S_0$, which is invariant under a symmetry group ${\cal G}$ acting in
the standard way\cite{KnizhnikZamolodchikov1984,Witten1984}. The chiral
components of the conserved Noether currents, $J^a(z)$ and ${\bar
J}^a({\bar z})$, depend (as indicated) only on $z=x+iy$ and ${\bar
z}=x-iy$, and satisfy the defining operator product expansion (OPE) of
the Affine Lie algebra (Kac-Moody algebra) at level $k$
\begin{equation}
\label{KacMoodyOPE}
J^a(z) J^b(0)
=
{ {k\over 2} \delta^{ab}
\over z^2} + {1\over z} f^{ab}_c J^c(0) + \dotsb \ ,
\qquad
{\bar J}^a({\bar z}) {\bar J}^b(0)
=
{ {k\over 2} \delta^{ab}
\over {\bar z}^2} + {1\over {\bar z}} f^{ab}_c {\bar J}^c(0) + \dotsb \ ,
\end{equation}
where $f^{ab}_c$ are the structure constants of ${\cal G}$.  (Repeated
indices are summed throughout this paper, unless stated otherwise.)
The model we study is defined by the action\footnote{\label{Zfoot}The
partition function is ${\cal Z} (g_{0}) = \int {\cal D}[{\rm fields}]
\ \exp ({-\cal S})$.}
\begin{equation}
\label{actionNATM}
{\cal S}= {\cal S}_0 + g_{0} \int_z  \Phi(z,{\bar z}),
\qquad \qquad
\int_z := \int {\rmd^2 z\over 2\pi}\ ,
\end{equation}
where the perturbing operator
\begin{equation}
\label{DefPhi} 
\Phi(z,{\bar z}) \equiv J^a(z) {\bar J}^a({\bar z})
\end{equation}
is invariant under global transformations of the symmetry group ${\cal
G}$.  This theory is known to be renormalizable with a single coupling
$g$.  The conjectured form\cite{GerganovLeClairMoriconi2001} of the
$\beta$-function for the renormalized coupling $g$ is quoted in
(\ref{conjecturedbetafct}).

We compute the $\beta$-function explicitly to 4-loop order.  To this
end, consider the perturbative evaluation of the expectation value in
the fully interacting theory of some quantity ${\cal O}$, which may
represent an operator, or a product of operators at different spatial
positions,
\begin{equation}
\label{ExpectationValueO} 
\left<{\cal O}\right>_{g_0}={ {\cal Z}(0) \over
{\cal Z}(g_0) } \left< {\cal O} \ \rme^{- g_0 \int_z \Phi(z,{\bar
z})}\right>_0 \ .
\end{equation}
Here, the expectation value $\left< \dotsb \right>_0$ is taken in the
unperturbed CFT with action ${\cal S}_0$, normalized such that $\left<
1 \right>_{0}=1$. ${\cal Z}(g_0)$ is the fully interacting partition
function$^{\mbox{\scriptsize\ref{Zfoot}}}$.  A cut-off
(regularization) procedure, depending on short- and large-distance
cut-offs $a$ and $L$, is required to render all terms in this
expansion finite, and is specified below.  The renormalized coupling
$g$ (which depends on $g_0$ and ${a\over L}$) is found by computing
how the coefficient $g_0$ of the first order term in the expansion of
the exponential
\begin{eqnarray}
\label{ExpansionExponential} \nonumber \left< \dotsb ~ 1 \right>_0 &+&
\left< \dotsb \biggl ( - g_0 \int_{z_1} \Phi(z_1,{\bar z}_1) \biggr
)\right>_{\!\!0} + \left< \dotsb   \biggl ( { g_0^2\over 2!}  \int_{z_1} \int_{z_2}
\Phi(z_1,{\bar z}_1) \Phi(z_2,{\bar z}_2)
 \biggr )\right>_{\!\!0} \\
&+& \left< \dotsb \biggl ( {- g_0^3\over 3!}  \int_{z_1} \int_{z_2}
\int_{z_3} \Phi(z_1,{\bar z}_1) \Phi(z_2,{\bar z}_2) \Phi(z_3,{\bar
z}_3) \biggr )\right>_{\!\!0} + \dotsb
\end{eqnarray}
is modified by the higher order expansion terms.  This modification is
independent of the potential presence of any operator ${\cal O}$ in
this expectation value, indicated by the ellipses\footnote{The
operator itself requires an analogous treatment, which, however, can
be discussed independently; this will not be needed here.}.  The
required calculation can be conveniently expressed in terms of
multiple OPE coefficients of the perturbing operator $\Phi(z,{\bar
z})$, evaluated in the unperturbed theory.  The product of $(n+1)$
such operators at different positions may be expanded into a ``complete
set'' of operators $\Phi^{A}$ sitting at the position of, say, the
last operator.  The expansion coefficients depend on the $n$ relative
coordinates,
\begin{eqnarray}
\label{DefMultipleOPEcoeffC}
&&\hspace{-.2cm}
\Phi (z_{1},\bar z_{1})\dots \Phi (z_{n},\bar z_{n})\Phi (z_{n+1},\bar z_{n+1}) = \\
&&=\sum_{A} {\bf C}_A
\bigl [
(z_1-z_{n+1}),({\bar z}_1-{\bar z}_{n+1});
 \dotsb  ;(z_n-z_{n+1}),({\bar z}_n-{\bar z}_{n+1})
\bigr ]\,
\Phi^{A}(z_{n+1},\bar z_{n+1}) \ .\nonumber
\end{eqnarray}
The non-vanishing expansion coefficients are exactly known in any CFT.
In the present case they are especially simple, and can be obtained by
successive use of the OPE of the currents, (\ref{KacMoodyOPE}).  In
particular, the perturbing operator $\Phi$ is the most relevant
operator (besides the identity when $k \not = 0$) appearing amongst
the $\Phi^{A}$; all others are irrelevant.  We find it convenient to
denote the needed {\it multiple OPE coefficient}, where
$\Phi^{A}=\Phi$, by the symbol
\begin{eqnarray}
\label{DefPhiPhi|Phi}
\nonumber
 \MOPE{\,\rule{0mm}{2ex}\Phi (z_{1},\bar
z_{1})\dots \Phi (z_{n},\bar z_{n})\Phi (z_{n+1},\bar z_{n+1})
\,}{\,\Phi(z_{n+1},\bar z_{n+1})\,}\qquad \qquad \qquad 
\\
= {\bf C}_\Phi \bigl[ (z_1-z_{n+1}),({\bar z}_1-{\bar z}_{n+1});
\dotsb ;(z_n-z_{n+1}),({\bar z}_n-{\bar z}_{n+1}) \bigr ]\ .
\end{eqnarray}
The renormalization process is now easily understood by inserting
(\ref{DefMultipleOPEcoeffC}) into
(\ref{ExpansionExponential}). Explicitly, denote the relevant
integrals over the multiple OPE coefficients by ${\cal F}_{n}$ (for
``Feynman''-diagram):
\begin{equation}
\label{integralsMOPE} {\cal F}_{n} :=\!\int _{z_{1},z_{2}, \dots z_{n}}\!\!
\MOPE{\,\rule{0mm}{2ex}\Phi (z_{1},\bar z_{1})\dots \Phi (z_{n+1},\bar
z_{n+1}) \,}{\,\Phi (z_{n+1},\bar z_{n+1})\,} \ {\cal C} (z_{1},\bar
z_{1}, \dots, z_{n+1}, \bar z_{n+1})\ .
\end{equation}
These integrals are regularized by a cut-off prescription,
which is achieved by inserting a cut-off function \\
$ {\cal C}
(z_{1},\bar z_{1}, \dots, z_{n+1}, \bar z_{n+1}) $ in the integral, as
indicated. There are many possible choices.  In this paper we choose a
(circular) hard cut-off implemented by
\begin{equation}
\label{DEFcutoffFunction} \label{lf16} {\cal C} (z_{1},\bar z_{1},
\dots, z_{n}, \bar z_{n}) := \prod_{i\neq j} \Theta (a<|z_{i}-z_{j}|<L)
\ ,
\end{equation}
where $\Theta$ is the usual step function. This cut-off procedure
restricts the distances between any pair of integration variables to
lie between the short- and the long-distance cut-offs $a$ and $L$.
All integrals ${\cal F}_n$ are thus finite functions of ${a\over L}$.
As usual, inserting (\ref{DefMultipleOPEcoeffC}) in
(\ref{ExpansionExponential}), and using (\ref{integralsMOPE}) gives:
\begin{eqnarray}
\label{ExpansionMOPE} \nonumber \left< \dotsb 1 \right>_{0} &+& \left<
\dotsb \biggl ( - g_0 \int_{z} \Phi(z,{\bar z}) \biggr
)\right>_{\!\!0} + \left< \dotsb \biggl ( { g_0^2\over 2!}  {\cal F}_1
\int_{z} \Phi(z,{\bar z})
 \biggr )\right>_{\!\!0} \\
\nonumber &+& \left< \dotsb \biggl ( {- g_0^3\over 3!} {\cal F}_2
\int_{z} \Phi(z,{\bar z})
 \biggr ) \right>_{\!\!0} + \dotsb   \\
\nonumber & &=\left< \dotsb 1 \right>_0 + \left< \dotsb \biggl ( - g \int_{z}
\Phi(z,{\bar z}) \biggr )\right>_{\!\!0} +\dotsb =\left< \dotsb \biggl
(\rme^{- g \int_{z} \Phi(z,{\bar z})} \biggr ) \right>_{\!\!0}\ .
\end{eqnarray}
Following standard reasoning we have re-exponentiated in the last
line.  One can now read off the renormalized coupling:
\begin{equation}
\label{DEF g_R} g\! \left( g_0, {a\over L}\right) = g_0 \bigg[ 1
-{g_0\over 2!}  \,{\cal F}_1\! \left( {{a\over L}}\right) +{g_0^2
\over 3!} \,{\cal F}_{2}\!\left({a\over L} \right) - ...  \biggr ] \ .
\end{equation}
The $\beta$-function is obtained as the change of $g$ in response to
changing $a$ (or $1/L$), while keeping the bare coupling $g_0$ fixed:
\begin{equation}
\label{bd}
\beta(g) := a {\partial \over \partial  a}\bigg|_{ g_0 }
g\!\left( g_0, {a\over L}\right)\ .
\end{equation}
In the remaining sections of the paper we will obtain the integrals
${\cal F}_1, \dotsb , {\cal F}_4$. This gives us the result for the
4-loop $\beta$-function written in (\ref{beta-4-loop-unspec}) above.


\section{Calculation}
\label{secCalculation}
In this section we present in detail the evaluation of the integrals
${\cal F}_n$ defined in (\ref{integralsMOPE}) (``Feynman''-diagrams),
needed to obtain the $\beta$-function, as explained in Section
\ref{Model and Method}.  The core of this calculation consists in
obtaining the OPE coefficients defined in (\ref{DefPhiPhi|Phi}), by
repeated use of the current algebra OPE (\ref{KacMoodyOPE}).  We start
with the simplest case, i.e.\ with the 1-loop integral ${\cal F}_1$,
and proceed successively to the more involved cases, up to 4-loop
order.

\subsection{1-loop order}\label{1loop}
At 1-loop order, we need the  OPE-coefficient 
\begin{equation}\label{k1}
\MOPE{\,\rule{0mm}{2ex}\Phi (z,\bar z)\Phi (w,\bar w)\,}{\,\Phi
(\bar w,w ) \,}\ .
\end{equation}
To evaluate it, we have to eliminate $\Phi (z,\bar z)$ from $\Phi
(z,\bar z)\Phi (w,\bar w)$, using (\ref{KacMoodyOPE}). This is done as
follows\footnote{Recall that summation over repeated indices is
implied, and that we work at level $k=0$.}
\begin{eqnarray}\label{lf18}
&& \Phi (z,\bar z) \Phi (w,\bar w)  \equiv  J^{a} (z) \bar J^{a} (\bar
z)  J^{b} (w) \bar J^{b} (\bar w)\nn \\
&&\qquad \qquad \longrightarrow 
f^{ab}_{c}f^{ab}_{d} J^{c} (w) \bar J^{d} (\bar w) \frac{1}{|w-z|^{2}} 
= C_{2} J^c (w)\bar J^{c} (\bar w)  \frac{1}{|w-z|^{2}}
\ .\qquad  \label{lf19}
\end{eqnarray}
We have used that $f^{abc}f^{abd}=C_{2}\delta^{cd}$, with the second
Casimir $C_{2}$. (This and more group theoretical relations are
derived in appendix \ref{a:algebra}.)  We denote (\ref{k1}) in short
by
\begin{equation}\label{lf20} 
\MOPE{\,\rule{0mm}{2ex}\Phi (z,\bar z)\Phi (w,\bar w)\,}{\,\Phi
(\bar w,w ) \,}= \ 
{}_{z}\diagram{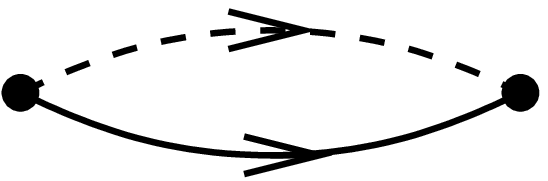}{}_{w} \ =\  {}_{z}\diagram{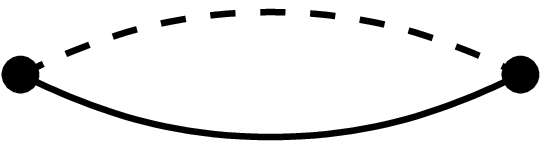}{}_{w} \ = \frac{
C_{2}}{|w-z|^{2}} \ .
\end{equation} 
The arrows show the direction in which the elimination has been
made. This defines the sign. To be specific, an arrow from $z$ to $w$
represents $1/ (z-w)$. A dashed  such arrow represents
$1/ (\bar z-\bar w)$. When a solid
and a dashed arrow (with the same direction)
 connect the same two points, one can drop the
arrows for simplicity of notation; seeing a solid and a dashed line
thus means that when adding the arrows, both arrows are pointing in
the same direction.  

The OPE-coefficient (\ref{lf20}) yields the 1-loop diagram ${\cal F}_{1}$
\begin{equation}\label{lf21}
{\cal F}_{1}=\int_{z}\diagram{1ban}\nn \\
=\int_{z} \frac{ C_{2} }{|w-z|^{2}}  \Theta
(a<|w-z|<L) =  C_2 \,\ln\! \Big( \frac{a}{L}\Big)
\ . \qquad 
\end{equation}

\subsection{2-loop order and the magic rule}\label{2loop}
\begin{figure}
\centerline{\fig{.65\columnwidth}{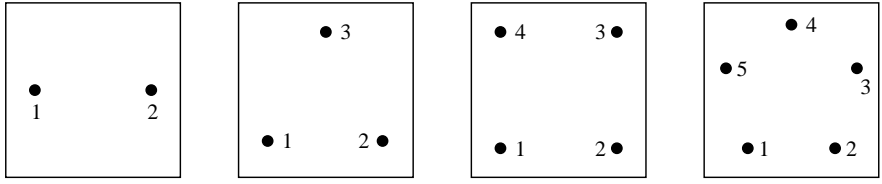}}
\caption{Labeling of the points in figures \ref{2loopalldiags},
\ref{3loopreddiags}, \ref{f:4loop-remain}, and diagrams in the main text.}
\label{f:labelpoints}
\end{figure}%
\begin{figure}[htb]
\centerline{\fig{0.65\columnwidth}{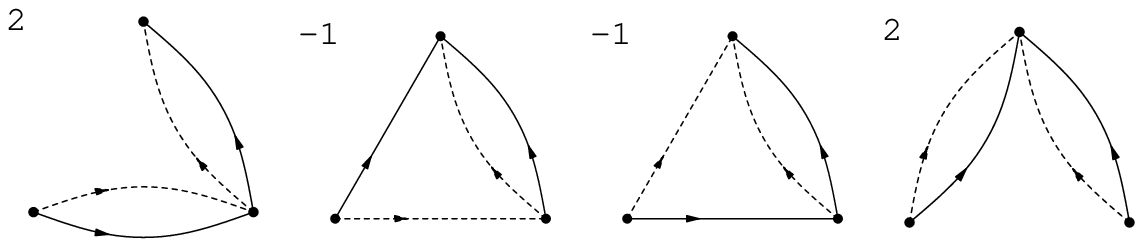}}
\medskip 
\centerline{\hphantom{ magic rule}\parbox{.2in}{{\Huge$ \downarrow$}}
magic rule} 
\medskip 
\centerline{\fig{0.4875\columnwidth}{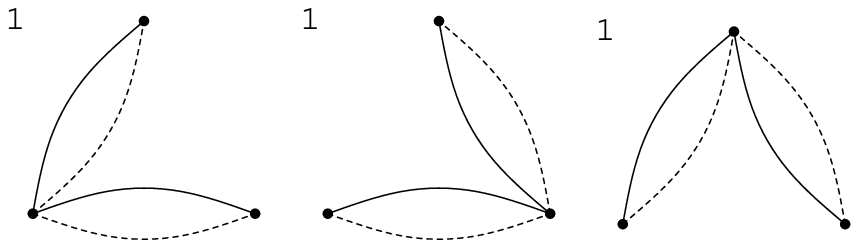}}
\caption{2-loop diagrams after reducing the structure-constants to
numbers. To be multiplied by $\half C_2^2 $. The numbers
given are the weight. The first
line is the raw result, as obtained by using the reduction
algorithm. Arrows indicate the direction of the reduction. The second
line after using magic relations.} 
\label{2loopalldiags}
\end{figure}%
At 2-loop order, we have three $\Phi $'s. Denoting $\Phi
_{i}:=J^{a} (z_{i})\bar J^{a} (\bar z_{i})$, we need to calculate $
\MOPE{\,\rule{0mm}{1.8ex}\Phi_{1} \Phi _{2}\Phi _{3}\,}{\,\Phi_{3}
\,}$.  Straightforward use of
the current-algebra (\ref{KacMoodyOPE})
with
$k=0$,  eliminating the currents
one by one,  starting with point 1, and continuing with point 2, yields
\begin{equation}\label{2lope}
\MOPE{\,\rule{0mm}{2ex}\Phi_{1} \Phi _{2}\Phi _{3}\,}{\,\Phi_{3} \,}
= \frac{1}{2} C_{2}^{2}\, \bigg[
\frac{2}{|z_{12}|^{2}|z_{23}|^{2}} +\frac{2}{|z_{13}|^{2}|z_{23}|^{2}}
- \frac{1}{|z_{23}|^{2} z_{13}\bar z_{12}}- \frac{1}{|z_{23}|^{2}
z_{12}\bar z_{13}} 
 \bigg] 
\end{equation} 
 where we have abbreviated $z_{ij}:=z_{i}-z_{j}$.
Here, and throughout this article, we use the labeling of points as indicated in
figure \ref{f:labelpoints}.
The result is graphically presented in figure
\ref{2loopalldiags} (top). 
  Equation
(\ref{2lope}) apparently contains a new diagram, which renders the
OPE-coefficient asymmetric upon exchange of point 1 with point 2, or
of point 1 with point 3. However, there is a simple algebraic
identity, the ``magic'' rule for the real part $\Re $ of $\frac{1}{z\bar w}$:
\begin{equation}\label{lf23}
\Re \left[\frac{1}{z \bar w}  \right] = \Re \left[\frac{\bar z
w}{|z|^{2} |w|^{2}}   \right] =  \left[\frac{\vec{z}\vec{w} }{|\vec  z|^{2}
|\vec  w|^{2}}  \right] = \half \left[\frac{1}{|\vec  z|^{2}} + \frac{1}{|\vec w|^{2}}-\frac{|\vec{z}-\vec{w}  |^{2}}{|\vec  z|^{2}|\vec  w|^{2}} \right]
\ .
\end{equation}
The most useful application is in the presence of an additional
factor $1/|\vec{w}-\vec{z}| ^{2} $, which cancels the numerator in the
last term. This leads to the decomposition of the new diagrams in
(\ref{lf23}) into chain-diagrams [drawn below
rotated by $-120^0$ as compared
to figure \ref{2loopalldiags} (top)]
\begin{equation}\label{lf24}
\diagram{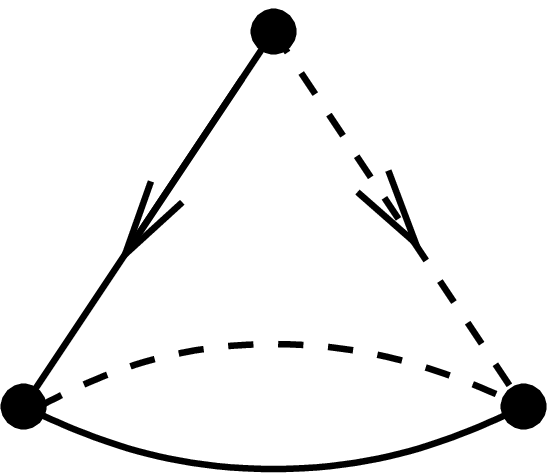}+\diagram{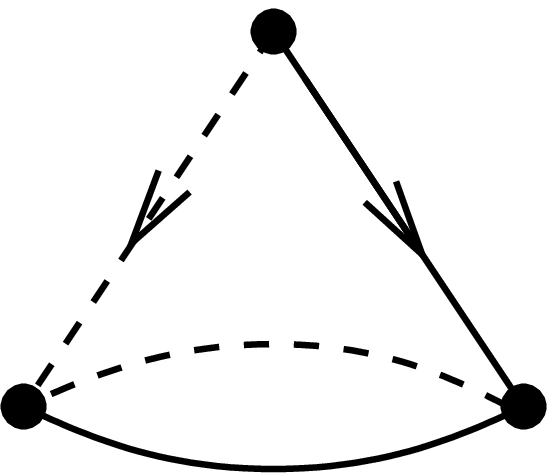}  = \diagram{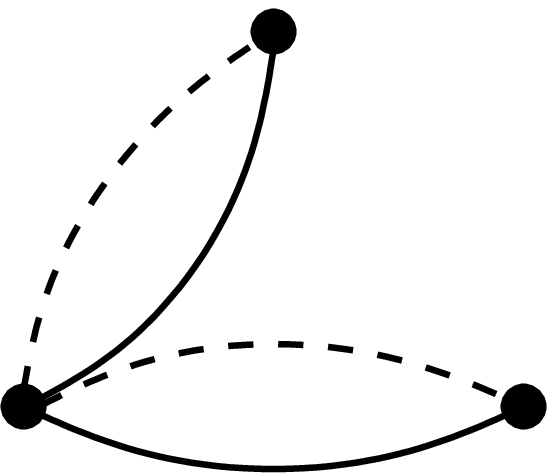}+
\diagram{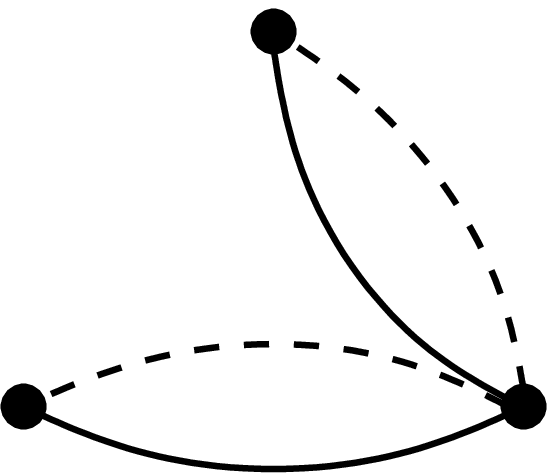} - \diagram{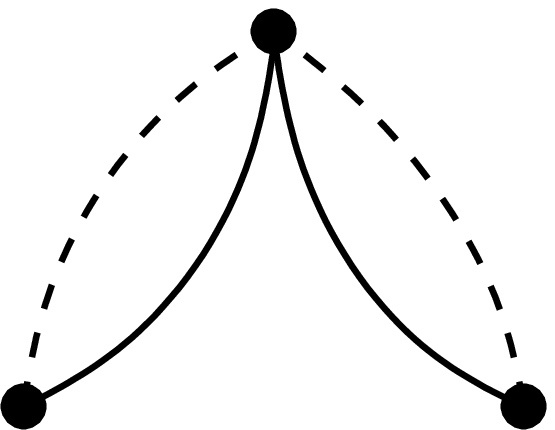} 
\ .
\end{equation}
The OPE-coefficient (\ref{2lope}) simplifies to 
\begin{equation}\label{t1}
\MOPE{\,\rule{0mm}{2ex}\Phi_{1} \Phi _{2}\Phi _{3}\,}{\,\Phi_{3} \,} =
\frac{1}{2} C_{2}^{2}\, \left[\frac{1}{|z_{12}|^{2}|z_{13}|^{2}} +
\frac{1}{|z_{12}|^{2}|z_{23}|^{2}}+ \frac{1}{|z_{13}|^{2}|z_{23}|^{2}}
\right] \ ,
\end{equation}
which is manifestly symmetric, as it should be.  The resulting
expression for the OPE coefficient in (\ref{t1}) is graphically
represented in figure \ref{2loopalldiags} (bottom).  One sees that
after using the ``magic'' rule, the OPE coefficient, and hence the
integral ${\cal F}_2$, can be written in terms of chain diagrams.
This suggests, that the corresponding diagrams (i.e.\ the ``Feynman''
integral ${\cal F}_2$) factorize, are of order $\ln (\frac{L}{a})^{2}$
without a pure $\ln (\frac{L}{a})$ and thus give no contribution to
the $\beta$-function at 2-loop order. This is indeed correct, as
checked in appendix \ref{app:2loopfac} for the cutoff-procedure
introduced in section \ref{Model and Method}.

For the model at hand, the cut-off procedure is subtle. The reason is
that one cannot put a cut-off on the lines, as would be most
convenient to immediately prove factorization of chain-diagrams: In
constructing the diagram, we have used magic rules to move around the
lines, and if we leave behind a cut-off function, then the resulting
diagram will not be totally symmetric, as it should and as it is in
our construction. The only way out of the above dilemma, is to put
cut-offs between any pair of points, regardless of whether the two
points are connected with a line or not [compare
(\ref{integralsMOPE})].  However then the factorization is no longer a
trivial statement, and has to be checked. This has been done for the
2-loop chains in appendix \ref{app:2loopfac}.  As we have argued in
section \ref{present results}, this is not essential for our
arguments, and the conclusions remain valid in any scheme.  Let us
however mention, that in order to recover the large-$N$ limit of
$\mbox{SU} (N)$, factorization is needed, and is sufficient to {\em
uniquely} fix the RG-procedure up to 4-loop order; but not necessarily beyond.

\begin{figure}[t]
\centerline{\fig{1.06\textwidth}{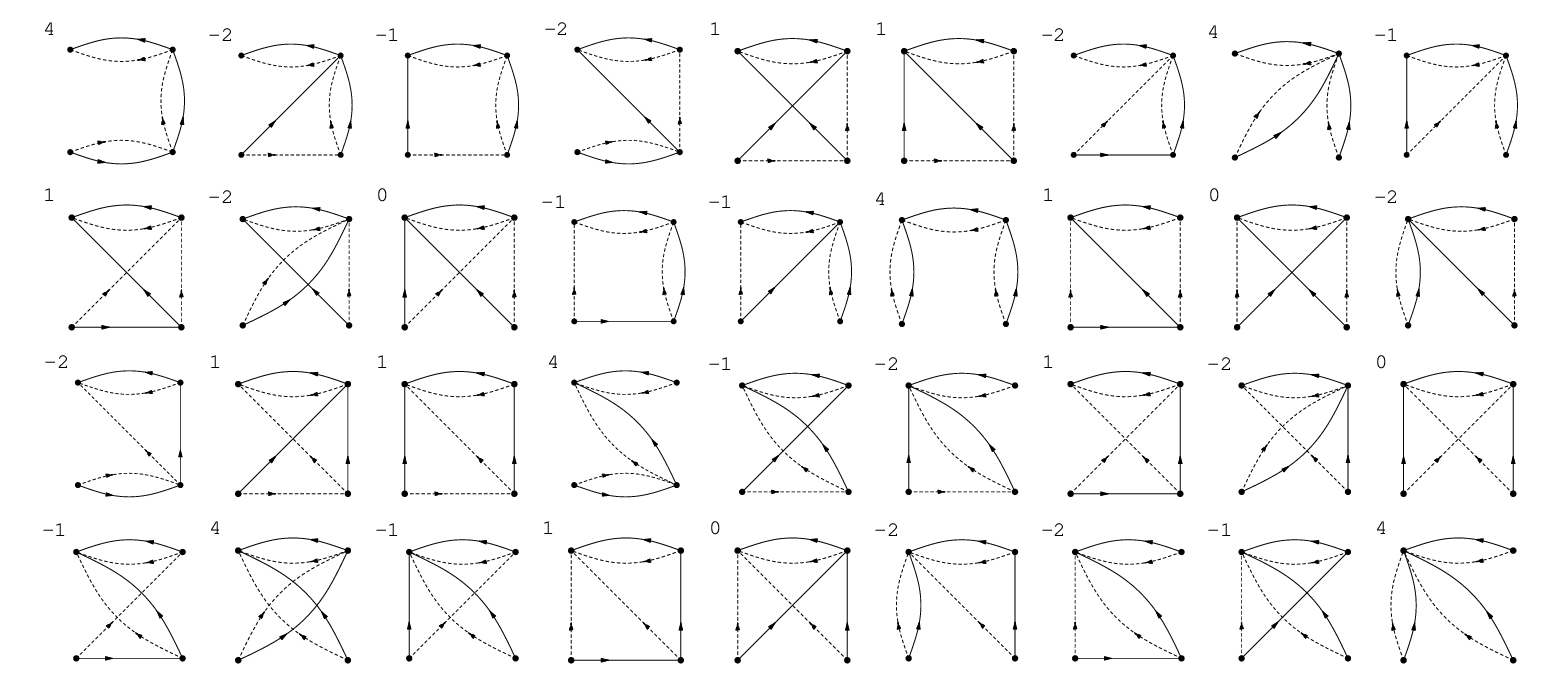}}
\medskip\centerline{\hphantom{ magic rule}\parbox{.2in}{{\Huge$ \downarrow$}}
magic rule}  \medskip 
\centerline{\fig{0.7\textwidth}{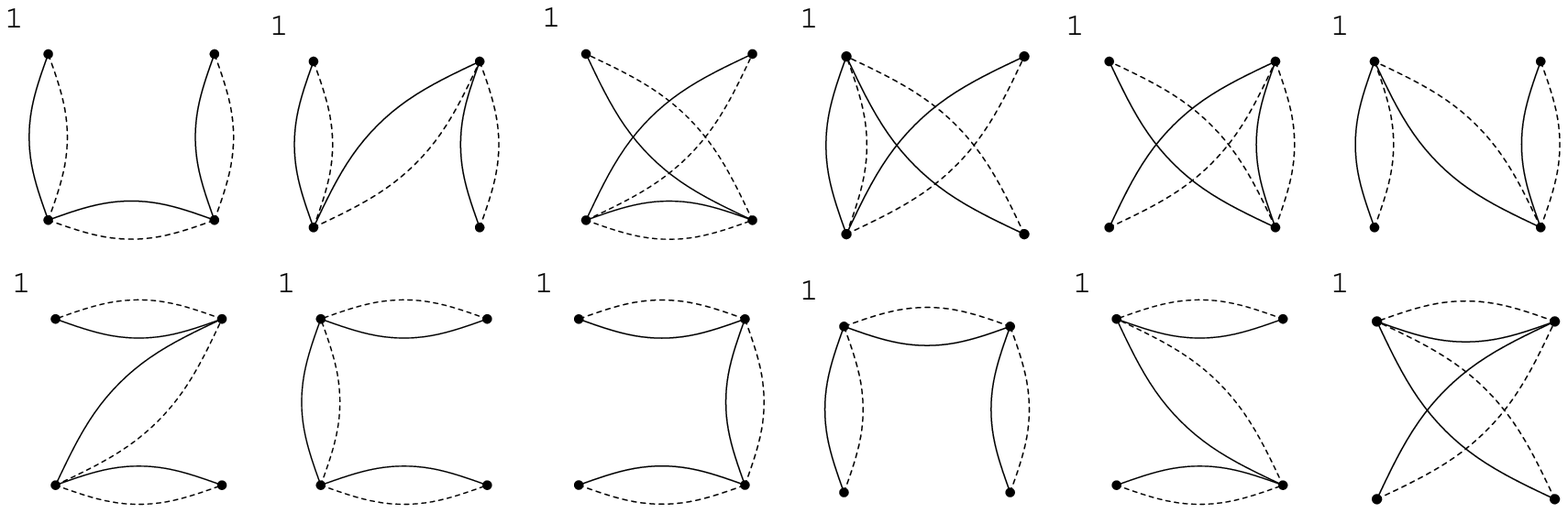}}
\caption{3-loop diagrams after reducing the structure-constants to
numbers. To be multiplied by $\frac{1}{4} C_2^3 $. The numbers
given are the weight. The first 4
lines are the raw result, as obtained by using the reduction
algorithm. Note that 4 diagrams have 
weight  0. Arrows indicate the direction of the
reduction. The last 
two lines after using magic relations,  dropping the redundant arrows.} 
\label{3loopreddiags}
\end{figure}%
\subsection{3-loop order}\label{3loop} At 3-loop order, 36
diagrams appear, presented on top of figure \ref{3loopreddiags}. These
diagrams all contain six structure-constants, and have two free
indices $a$ and $b$, which are contracted with the remaining $\bar
J^{a}J^{b}$. Since the only invariant object with two indices is
$\delta^{ab}$, one can contract the last lines to obtain the algebraic
factor; the final result has of course to be divided by the dimension
of the adjoint representation. One can then convince oneself by
drawing pictures, that all objects which can be constructed, contain
at least one loop made out of two or three vertices\footnote{The
simplest object without such a loop would be a cube, which indeed
appears at 4-loop order, see figure \ref{cube}.}, i.e.\ objects of the
form
\begin{eqnarray}\label{alg1}
f^{acb}f^{bcd}&=& \vardiagram{0.3}{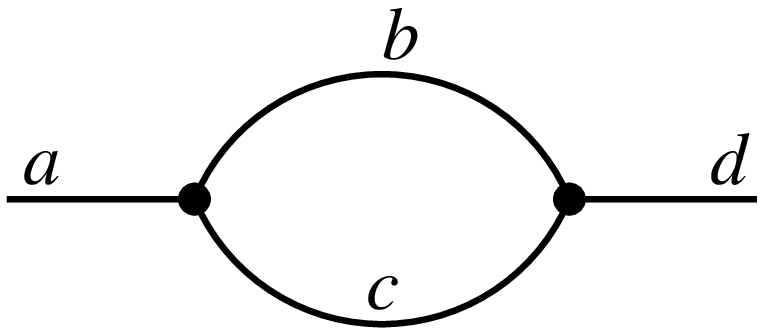} =- C_{2} \delta^{ad}\\
\label{alg2}
f^{aed}f^{bge}f^{cdg}&=&\vardiagram{0.3}{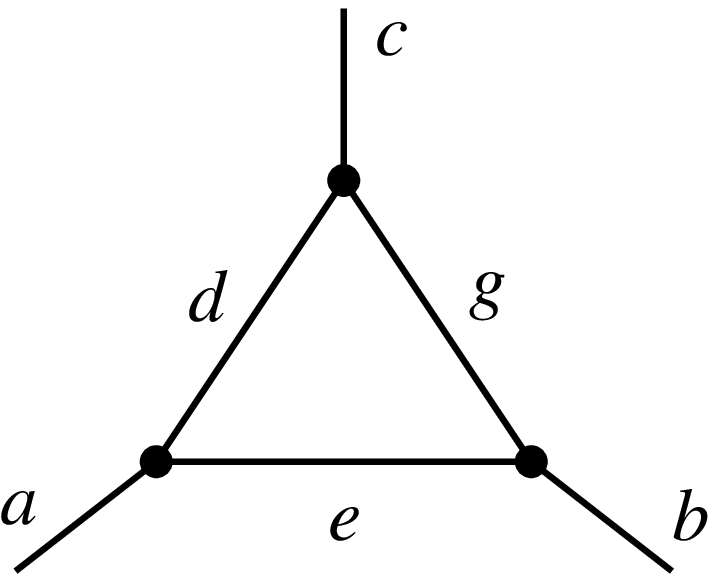} = -\frac{C_{2}}{2} f^{abc}\ ,
\end{eqnarray}
which we give together with a
group-theoretical identity (derived in appendix \ref{a:algebra}),
which is sufficient to reduce the number of $f$ in any given
diagram. Repeatedly using (\ref{alg1}) and (\ref{alg2}) thus allows to
eliminate all $f$. This procedure is performed using a computer, and
the reader would have a hard time verifying it by hand.  We have thus
shown that all 3-loop diagrams are proportional to $C_{2}^{3}$, thus
no additional group theory invariants, besides the second Casimir, appear
at this order.

The diagrams are given with their combinatorial factor on top of
figure \ref{3loopreddiags}, to be multiplied by $\frac{1}{4}
C_{2}^{3}$. Applying magic rules leads to chain-diagrams, presented
graphically at the bottom of figure
\ref{3loopreddiags}. Algebraically, the result is 
\begin{eqnarray}\label{lf7}
\MOPE{\Phi _{1}\Phi_{2}\Phi _{3}\Phi _{4}}{\Phi _{1}}
&=&\frac{C_{2}^{3} }{4}\Bigg[
\frac{1}{|z_{12}|^2|z_{14}|^2|z_{23}|^2}
+\frac{1}{|z_{13}|^2|z_{14}|^2|z_{23}|^2}
+\frac{1}{|z_{12}|^2|z_{13}|^2|z_{24}|^2}\nn \\
&&\hphantom{\frac{C_{2}^{3} }{4}\Bigg[}
+\frac{1}{|z_{13}|^2|z_{14}|^2|z_{24}|^2}
+\frac{1}{|z_{13}|^2|z_{23}|^2|z_{24}|^2} 
+\frac{1}{|z_{14}|^2|z_{23}|^2|z_{24}|^2}\nn \\
&&\hphantom{\frac{C_{2}^{3} }{4}\Bigg[}
+\frac{1}{|z_{12}|^2|z_{13}|^2|z_{34}|^2}
+\frac{1}{|z_{12}|^2|z_{14}|^2|z_{34}|^2}
+\frac{1}{|z_{12}|^2|z_{23}|^2|z_{34}|^2}\nn \\
&&\hphantom{\frac{C_{2}^{3} }{4}\Bigg[}
+\frac{1}{|z_{14}|^2|z_{23}|^2|z_{34}|^2} 
+\frac{1}{|z_{12}|^2|z_{24}|^2|z_{34}|^2}
+\frac{1}{|z_{13}|^2|z_{24}|^2|z_{34}|^2} \Bigg]\!\!\nn \\\label{lf25}
\end{eqnarray}

\begin{figure}[t]
\centerline{\fig{1.055\textwidth}{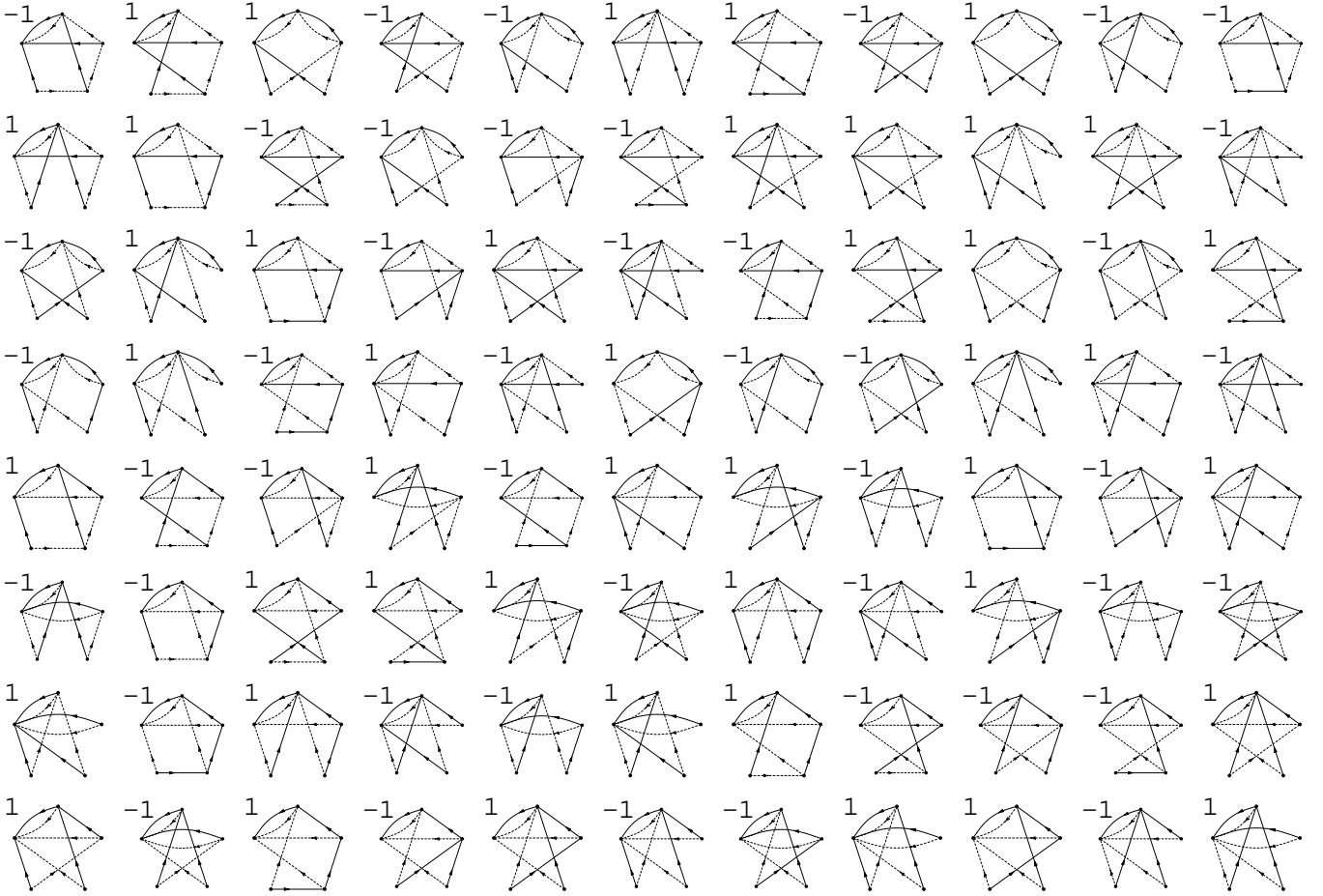}}
\caption{4-loop diagrams not proportional to $C_{2}^{4}$.}
\label{f:4loop-remain}
\end{figure}%
\subsection{4-loop order, direct approach}\label{4loop direct} Let us
now continue to 4-loop order. After using the current-algebra, there
are 576 diagrams, which again we generate computer-algebraically.  The
group theoretical factors appearing with these diagrams are much more
involved.  An example is a cube, where each corner represents a
structure-factor $f^{abc}$ and each link identifies a pair of common
indices between two $f$'s.  This is drawn on figure \ref{cube} and
detailed in appendix \ref{a:algebra}.  It is at this loop-order
that an additional group theoretical invariant besides the quadratic
Casimir arises.  After reducing the algebra, one finds that 380
terms are proportional to $C_{2}^{4}$. Using magic rules, these
diagrams can be reduced to 60 chains; these are in fact all the chains
which can be drawn through 5 points\footnote{ In general, there is a
total of ${1\over 2} n \cdot (n-1)!$ chains that can be drawn through
$n$ points.}. Each chain comes with a weight of $\frac{1}{8}C_{2}^{4}
$.
For the remaining diagrams not proportional to
$C_{2}^{4}$, presented in figure \ref{f:4loop-remain}, our
reduction-algorithm based on magic rules is incapable of further
simplifying it. In section \ref{4loop simplified} we will present a
simple calculation, reducing the task to calculating a combination of
eight diagrams. To this aim,  we need correlation functions
involving operators which we call
``adjoint'' perturbations, defined below.

\subsection{OPE for adjoint perturbations}\label{axialOPEs}
Define the ``adjoint'' perturbation at position  $(z_i, {\bar z}_i)$
as 
\begin{eqnarray}\label{lf8}
\Phi^{a} (z_{i}, \bar z_{i})\equiv \Phi^{a}_{i}:= f^{abc} J^{b}
(z_{i})\bar J^{c} (\bar z_{i}) \ .
\end{eqnarray}
We now apply the same procedure as in the previous sections:
Eliminate currents one by one using i)  the current-algebra, ii)
 evaluation of
the group theoretical factors, and  iii) simplifications with the magic rule. 
After some lengthy calculations 
(done again computer-algebraically), we find up to 3-loop order:
\begin{eqnarray}\label{lf9}
\MOPE{\rule{0mm}{2ex}\,\Phi _{1}^{a}\Phi_{2}^{a}\,}{\,\Phi_{2}\, } &=&
\frac{C_2^{2}}{2} 
\frac{1}{|z_{12}|^{2}} \\
\MOPE{\,\rule{0mm}{2ex}\Phi _{1}^{a}\Phi _{2}\Phi _{3}^{a}\,
}{\,\Phi_{3}\, } &=& \frac{C_{2}^{3}}{4}\left[ \frac{1}{|z_{12}|^2
|z_{13}|^2}+\frac{1}{|z_{13}|^2
|z_{23}|^2}\right]\label{lf9a}  \\
\MOPE{\,\rule{0mm}{2ex}\Phi _{1}^{a}\Phi _{2}^{a}\Phi _{3}\Phi _{4}
\,}{\,\Phi_{4}\, } &=& 
\frac{C_{2}^{4}}{8} \Bigg[
\frac{1}{|z_{12}|^2 |z_{14}|^2|z_{23}|^2}+\frac{1}{|z_{12}|^2|z_{13}|^2|z_{24}|^2}+\frac{1}{|z_{12}|^2|z_{13}|^2|z_{3
  4}|^2}\nn \\
&&\hphantom{
\frac{\, C_{2}^{4}}{8}}
+\frac{1}{|z_{12}|^2|z_{14}|^2|z_{34}|^2}
+\frac{1}{|z_{12}|^2|z_{23}|^2|z_{34}|^2} 
+\frac{1}{|z_{12}|^2|z_{24}   
  |^2|z_{34}|^2}  
\Bigg]\nn \\
&&
+ 
d_{2} 
\left[\diagram{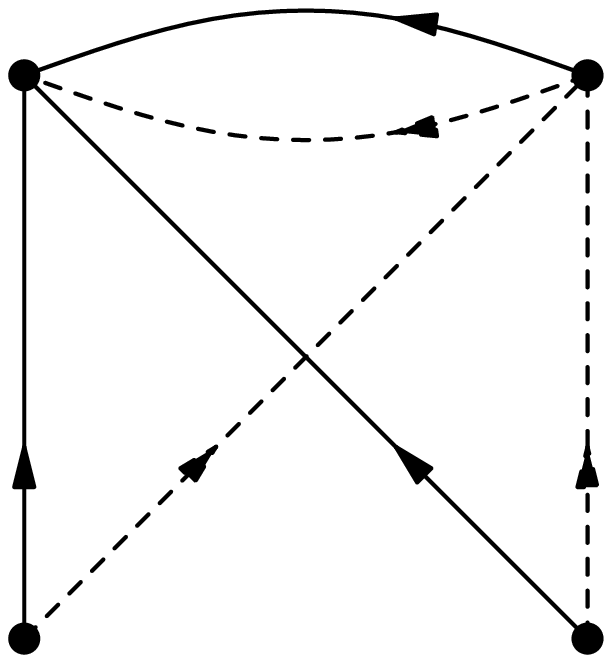}-\diagram{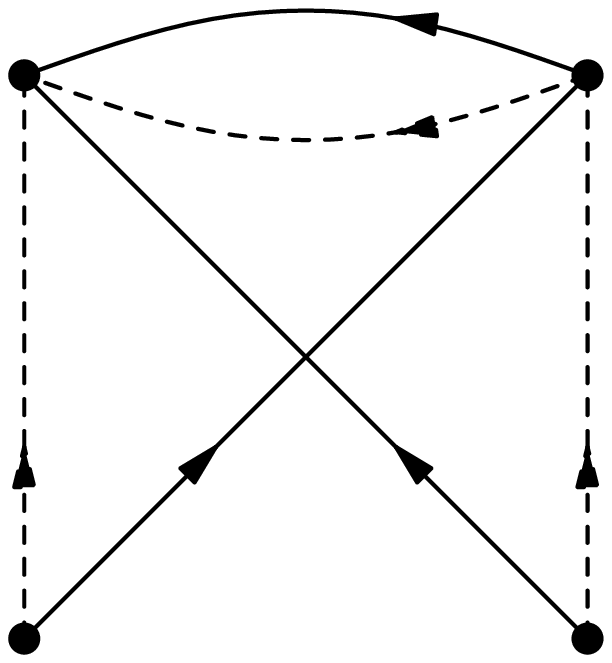}-\diagram{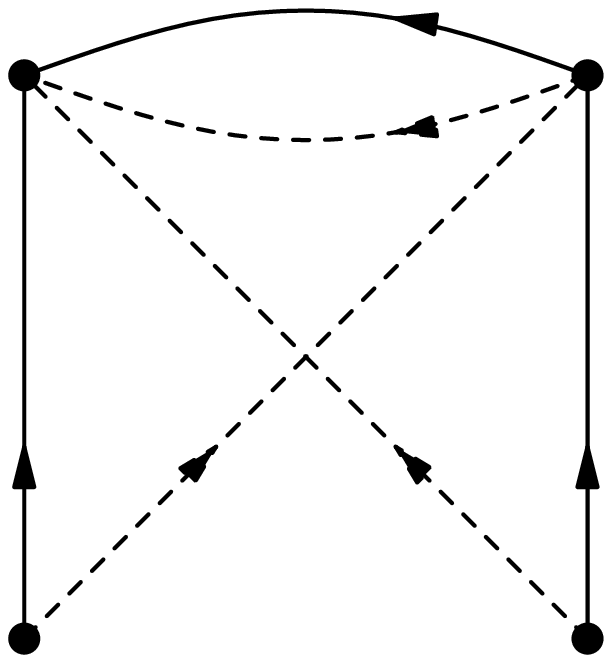}+\diagram{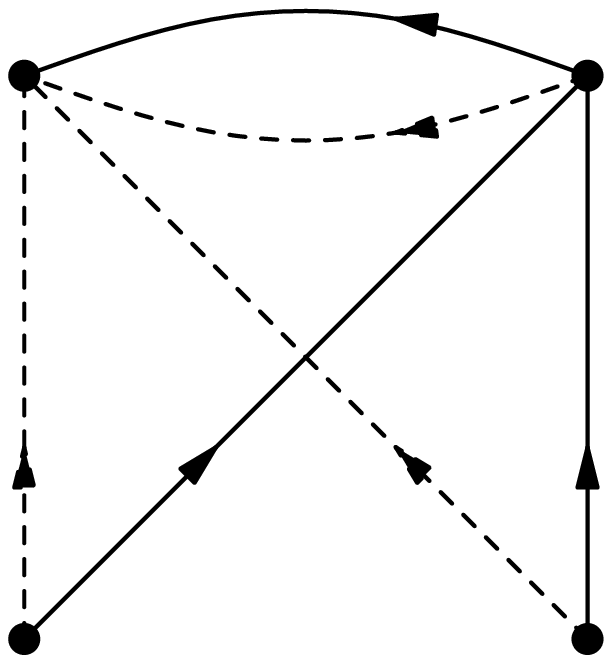}
\right]\!\!\!  \nn \\ \label{x1}
\end{eqnarray}
The additional group-theoretical invariant $d_{2}$ is defined
\footnote{Constructing a symmetrized tensor $d^{abcd}$ out of the
trace of 4 structure-constants $f^{ab}_{c}$, $d_{2}$ is the
non-dominant contribution (in $N$) of the square of $d^{abcd}$ [as
defined in (\ref{lf57})].}  in appendix \ref{a:algebra}.  For
$\mbox{SU} (N)$ this reads $d_{2}=\frac{3}{2}N^{2}$ compared to the
leading term $C_{2}^{4}=N^{2}$.  The results for $\mbox{SO} (N)$ and
$\mbox{SP} (N)$ are listed in figure \ref{grouptable}, see also
appendix \ref{otheralgebras}.  For these groups, $C_{2}^{4} \sim
N^{4}$ and again $d_{2}$ is subdominant, with $d_{2}\sim N^{3}$.

\subsection{4-loop order simplified}\label{4loop simplified}
We have seen in section \ref{4loop direct} that a direct 4-loop
calculation is quite cumbersome. Instead, we use here
 a different approach,
inspired by the original work by Kutasov \cite{Kutasov1989} . We
start by eliminating $\Phi _{n}$
from the multiple OPE coefficient (\ref{DefPhiPhi|Phi}).
 Let us first give the result and
then explain how we have obtained it:
\begin{eqnarray}
\MOPE{\,\Phi_{n}\Phi _{1}\Phi_{2}\dots \Phi _{n-1}\,}{\,\Phi _{n-1}\,}
&=&\nn \\
&&\hspace{-2cm}- \sum _{i,j=1\dots n-1, i\neq j}
\frac{1 }{z_{n}-z_{i}}\frac{1 }{\bar z_{n}-\bar z_{j}}
  \MOPE{\,\Phi_{1}\dots \Phi
_{i}^{a}\dots \Phi _{j} ^{a} \dots  \Phi
_{n-1}\,}{\,\Phi _{n-1}\,} \nn \\
&&\hspace{-2cm}+\sum _{i=1}^{n-1} \frac{C_{2}}{|z_{n}-z_{i}|^{2}}
\MOPE{\,\Phi_{1} \dots  \Phi 
_{n-1}\,}{\,\Phi _{n-1}\,}\ .
\label{y0}
\end{eqnarray}
We have eliminated all currents at point $n$. Using the
current-algebra (\ref{KacMoodyOPE}) again with $k=0$, there is a
contribution from each pair of points $\left\{i,j \right\}$ with
$i,j\neq n $. The first line of (\ref{y0}) contains the contributions
with $i\neq j$, for which we have listed
below the corresponding
current-algebra identities in (\ref{y1}) and (\ref{y2}). The last line
of (\ref{y0}) is the case $i=j$, and is obtained by using the
current-algebra both for the holomorphic and antiholomorphic current, as
given in (\ref{y3}) below.
\begin{eqnarray}\label{y1}
J^{a} (z_n) \ \bar J^{b} (\bar z_{i}) J ^{b} (z_{i}) \ &\longrightarrow& \
\frac{ f^{abc}}{z_{n}-z_{i}}
J^{c} (z_{i}) \bar J^{b} (\bar z_{i}) = -\frac{1 }{z_{n}-z_{i}}\,\Phi
_{i}^{a}\\ 
\label{y2} \bar J^{a} (\bar z_{n})\ \bar J^{b} (\bar z_{j}) J ^{b} (z_{j})
\ &\longrightarrow& \ \frac{ f^{abc}}{\bar z_{n}-\bar z_{j}} \bar
J^{c} (\bar z_{j}) J^{b} (z_{j}) = \frac{1}{\bar z_{n}-\bar
z_{j}}\, \Phi _{j}^{a} \\ 
\label{y3} \bar J^{a} (\bar z_{n}) J^{a} (z_{n})\ \bar J^{b} (\bar
z_{i}) J^{b} (z_{i}) \ &\longrightarrow& \
\frac{f^{abc}f^{abd}}{|z_{n}-z_{i}|^{2}}\, J^{c} (z_{i}) \bar J^{d}
(\bar z_{i}) = \frac{C_{2}}{|z_{n}-z_{i}|}\,\Phi _{i}\ .
\end{eqnarray}
Note that eliminating point $n$ (instead of point 1 as we were used to
do) is for later calculational (and representational) convenience only.

We now turn to the 4-loop calculation, i.e.\ set $n=5$.  One can check
that starting from (\ref{y0}), using (\ref{lf25}) and (\ref{x1}), one
reconstructs all the 60 chains connecting 5 points, as found in
section \ref{4loop direct}. The remaining terms are obtained from the
first term in (\ref{y0}) times the term proportional to $d_{2}$ in
(\ref{x1}). There are $\frac{(n-1) (n-2)}{2}= 6$ such terms, each
being\begin{figure}
\centerline{\parbox{0.15\textwidth}{\fig{0.15\textwidth}{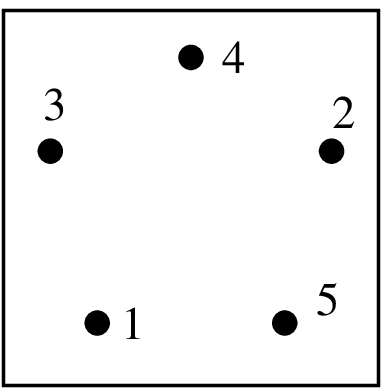}}\hfill
\parbox{.8\textwidth}{\fig{0.8\textwidth}{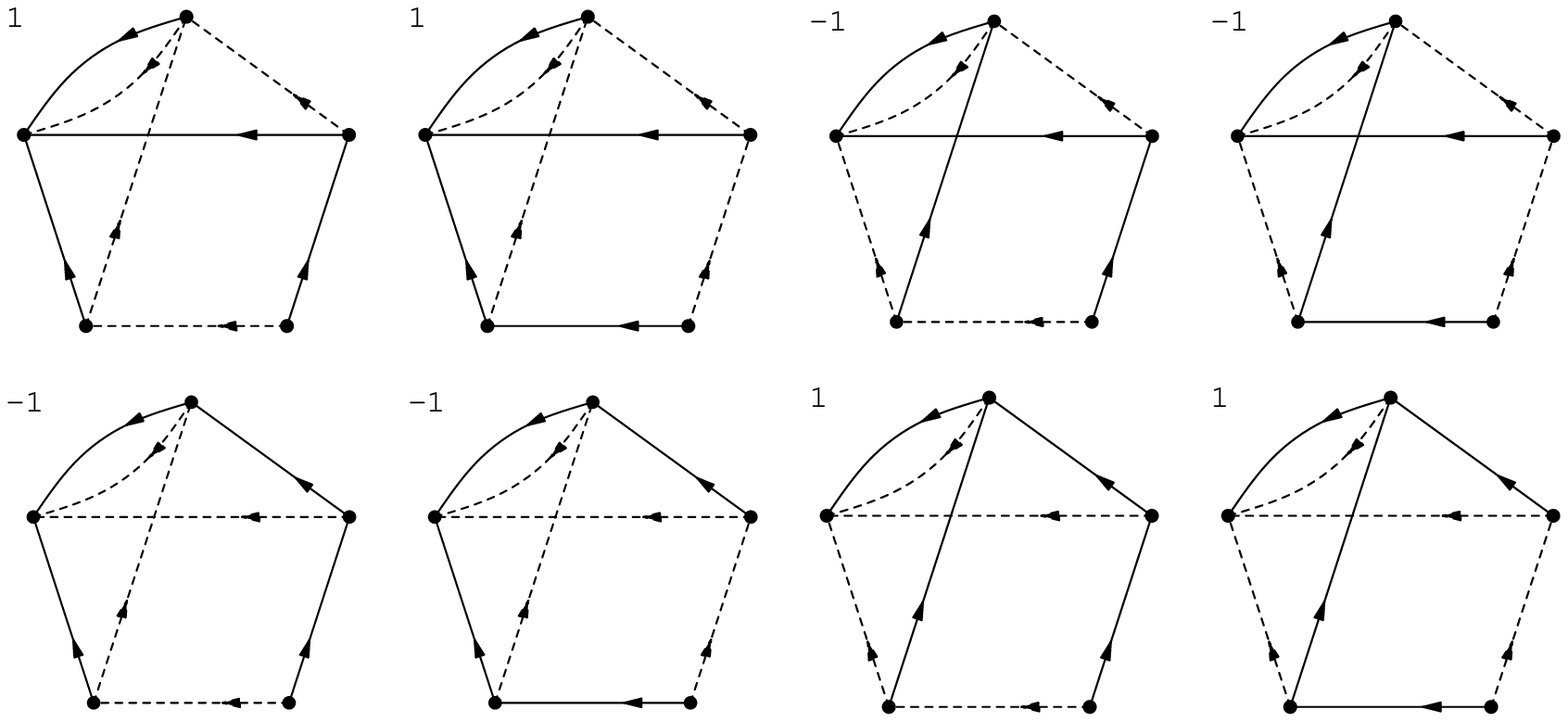}}} \caption{The
combination of 8 diagrams contributing at 4-loop order. Note the
different labeling of points given in the inset, as compared to the
labels of figure \ref{f:labelpoints}, used in figure
\ref{f:4loop-remain}.}  \label{f:4lMOPE}
\end{figure} a combination of 8 diagrams depicted in figure
\ref{f:4lMOPE}. Since each of the 6 terms gives the same contribution
upon integration, we only have to calculate the integral over one of
them. Analytically, this is most easily written as (we have chosen
$i,j$ to be the pair $1,2$ and the starting point is 5)
\begin{eqnarray}\label{y5}
{\cal I }&:=& \int \left(\frac{1}{z_{14}\bar z_{13}}- \frac{1}{\bar z_{14} z_{13}}\right)
\left(\frac{1}{z_{24}\bar z_{23}}- \frac{1}{\bar z_{24} z_{23}}\right)
\frac{1}{|z_{34}|^{2}} \left(\frac{1}{z_{15}\bar z_{25}}+
\frac{1}{\bar z_{15} z_{25}} \right) \nonumber \\
&&\qquad \times {\cal C} (z_{1},\bar z_{1},\dots
,z_{5},\bar z_{5})\ , 
\end{eqnarray}
where the integral is over all but one point, and the cutoff-function
${\cal C} (\dotsb)$ was introduced in (\ref{lf16}).  This integral is
evaluated in appendix \ref{4loopintegral} to be
\begin{equation}
\label{result 4loop integral}
{\cal I} = -\frac{\pi }{12} (6+\pi ^{2})\ln \left({ \frac{L}{a}}
\right) \ .
\end{equation}
\subsection{The $\beta $-function up to 4-loop order}\label{beta4loop}
Now we are ready to put everything together to obtain the $\beta$-function.
In a scheme in which the chains factorize, we obtain  by collecting the
results
 (\ref{lf21}), (\ref{t1}), (\ref{lf7}), and the paragraph  below (\ref{y0}),
and upon use of (\ref{DEF g_R})
:
\begin{eqnarray}\label{4l3}
g &=& g_{0} \Bigg\{ 1 - \frac{1}{2!} \left[g_{0} C_{2}\ln({\textstyle
\frac{L}{a}}) \right] + \frac{1}{3!} \frac{3}{2} \left[g_{0}
C_{2}\ln({\textstyle \frac{L}{a}} ) \right]^{2} -\frac{1}{4!}
\frac{12}{4} \left[g_{0} C_{2}\ln({\textstyle \frac{L}{a}}) \right]^{3}\nn \\
&&\hphantom{g_{0} \Bigg\{ 1} +\frac{1}{5!} \left( \frac{60}{8}
\left[g_{0}C_{2}\ln({\textstyle \frac{L}{a}} ) \right]^{4} +6
\frac{\pi }{12}\left(6+\pi ^{2} \right) d_{2}\, g_{0}^{4} \ln
({\textstyle \frac{L}{a}})\right)\Bigg\} \ .
\end{eqnarray}
Note that the factors $1/n!$ are from the exponential, then for the
chains the next factor is the number of chains times their dependence
on $g_{0}$, times the group-theoretical factor\footnote{See
(\ref{lf21}), (\ref{t1}), (\ref{lf7}), and the paragraph below
(\ref{y0}).} $ C_2^n/ 2^{n-1} $.  The last term (which comes from the
non-chain diagrams) has a factor of 6 from combinatorics as discussed
in the previous section and the minus signs from the integral
(\ref{result 4loop integral}) and from (\ref{y0}) cancel.

Inserting (\ref{4l3}) into (\ref{bd}) 
leads to the 4-loop $\beta$-function in terms of the
renormalized coupling $g$:
\begin{equation}\label{beta-4-loop}
\beta (g) = \frac{1}{2}  C_{2} g^{2} - d_{2} \frac{\pi }{240}
(6+\pi ^{2} )  g^{5} + O(g^6)\ ,
\end{equation}
with $d_2=\frac{3}{2}N^{2}$ for $\mbox{SU} (N)$,
$d_{2}=24-\frac{45}{2}N+6N^{2} -\frac{3}{8}N^{3}$ for $\mbox{SO} (N)$
and $d_{2}= \frac{3}{2}+\frac{45}{32}N +\frac{3}{8}N^{2}
+\frac{3}{128}N^{3}$ for $\mbox{SP} (N)$, as  calculated in appendices
\ref{SUn} and \ref{otheralgebras}.

In schemes in which the chains do not factorize, there are additional
terms, see (\ref{beta-4-loop-unspec}) and the discussion below that
equation.

Some comments on the procedure are in order. Readers used to the
Wilson-scheme, will recover that procedure by studying the change of
$g$ in (\ref{4l3}) under an infinitesimal change of $a$, corresponding
to the integration over an infinitesimal shell from $a$ to $a+\delta
a$. The only difference is that this is a shell in position space, and
not in momentum space.

Second, to our knowledge this is  the first 4-loop calculation with a
hard cutoff, or equivalently the first 4-loop calculation in a Wilson
scheme.


\section{Conclusion and further perspectives} \label{Conclusion} In
this article we have performed an explicit perturbative calculation of
the $\beta$-function for the non-abelian Thirring model at $k=0$ up to
4-loop order. We have found that the conjectured form of the
$\beta$-function \cite{GerganovLeClairMoriconi2001},
Eq.(\ref{conjecturedbetafct}), is incompatible with our result in {\em
all} regularization schemes.  The discrepancy arises from an extra
logarithmic divergence, which appears first at 4-loop order, and
which is proportional to a higher group-theoretical invariant
(evaluated in the adjoint representation of the symmetry group) which
is different from the quadratic Casimir invariant. This divergence is
not accounted for by the conjectured $\beta$-function.

It is worth pointing out that our explicit 4-loop result at level
$k=0$ does not only rule out the particular conjectured form of the
(isotropic) $\beta$-function Eq.~(\ref{conjecturedbetafct}), but a
more general class of conjectures for the $\beta$-function. This way
of presenting our 4-loop result emphasizes the dependence on the level
$k$, whereas in section \ref{present results} only the special case
$k=0$ was discussed.  Such forms which we can rule out arise by
attempting to ``scale'' with the level $k$. Specifically, for any one
of the classical groups ${\cal G} = \mbox{SU}(N), \mbox{SO}(N)$ and
$\mbox{SP}(N)$, the $\beta$-function (of the isotropic theory) will in
general be a function of three variables, the coupling constant $g$,
the level $k$, as well as $N$, or equivalently $C_2=C_2(N)$, the
second Casimir invariant in the adjoint representation:
\begin{equation}
\label{BetafctGeneralForm}
{\rmd g \over \rmd l}  = \beta ( g, k, C_2)\ .
\end{equation}
It was argued in \cite{BernardLeClair2001} that by rescaling the
Kac-Moody currents $J^a\to \sqrt{k} J^a$ (as suggested by the
large-$k$ calculations done in \cite{Kutasov1989}), the (isotropic)
$\beta$-function should satisfy  (perhaps in a suitable scheme)
 the ``scaling form''
\begin{equation}
\label{scalingform}
 \beta \left( g, k, C_2\right) \stackrel{\mbox{!}}{=} {1\over k} \,
 F\! \left( {C_2\over k}, k g\right)
= g  \,  H\! \left( g C_2,  {k \over C_2}\right)
\end{equation}
The conjectured $\beta$-function of Ref. \cite{GerganovLeClairMoriconi2001},
i.e.\ Eq.(\ref{conjecturedbetafct}), 
is a special case of this.  The second equation above gives an equivalent
way of writing the scaling form, useful when considering the limit
$k\to 0 $ for $g= \mbox{fixed}$, whereas the form in the first
equation is useful in the large $k$ limit where $1/k \to 0$ for
$kg=\mbox{fixed}$.  Since we know that the perturbative
$\beta$-function must have a finite limit as $k \to 0$, the second
equation in (\ref{scalingform}), when specialized to $k=0$, leads to a
form of the $\beta$-function, whose $g$-dependence is only through the
combination $g C_2$ (apart from an overall factor of $g$).  Comparison
with Eq.~(\ref{beta-4-loop-unspec}) shows that this is incompatible
with the explicit 4-loop result that we have found in any possible
scheme.  Hence, our result implies that the $\beta$-function must have
an explicit dependence on the level $k$, and that the latter can in no
scheme be ``scaled out'' in the way indicated in (\ref{scalingform}).

Finally, a point which deserves clarification is why in the case where
a symmetry ${\cal G}=\mbox{SU}(2)$ is broken down to $\mbox{U}(1)$ by
a purely imaginary easy-axis anisotropy, and for level $k=1$, the
conjecture reproduces \cite{BernardLeClair2001} known exact results
\cite{Nienhuis1984,FendleySaleurZamolodchikov1993a,FendleySaleurZamolodchikov1993b}.
(This was mentioned in the introduction, section \ref{intro}.)
Indeed, Ref. \cite{BernardLeClair2001} proposes that this agreement
should provide a strong check of the conjecture. Here we would like to
point out, however, that this agreement is not surprising, because
this theory is very special, and highly constraint by its underlying
hidden quantum group symmetry (or, ``fractional
supersymmetry'')\cite{BernardLeClair1990}, present for all values of
the level $k$.  The symmetry imposes strong constraints on the
$k$-dependence of the relationship between the slopes of the
$\beta$-function, i.e.\ the RG eigenvalues $y$ of the perturbation, at
the UV and IR fixed points.  As a consequence of the symmetry, this
relationship is \cite{LudwigWieseUnpublishedProgress}: 
\begin{equation}\label{gammarel}
\frac{1}{k\,y_{\mathrm{IR}}}+ \frac{1}{k\,y_{\mathrm{UV}}}=1\ .
\end{equation}
Note that the same result can also be obtained by using the
conjectured $\beta$-function of \cite{GerganovLeClairMoriconi2001},
see \cite{BernardLeClair2001}.  Now since, following Kutasov
\cite{Kutasov1989}, the conjecture is the leading term in a
$1/k$-expansion, it should yield a relation between
${y_{\mathrm{IR}}}$ and ${y_{\mathrm{UV}}}$, which is valid at leading
order in $1/k$, but will not contain information about corrections to
this of order $1/k^{2}$ or higher. However, due to the fractional
supersymmetry, the {\it exact} relation between ${y_{\mathrm{IR}}}$
and ${y_{\mathrm{UV}}}$ has no such higher order corrections at all,
as can be seen from Eq.~(\ref{gammarel}).  Thus, due to the
constraints imposed by this symmetry, the leading order term in $1/k$
happens to give already the whole, i.e.\ the exact result.  This
explains why the conjecture reproduces the exact result even for level
$k=1$.

We end our discussion by
noting that it would be interesting to obtain, generalizing
Kutasov's work, who computed (as mentioned) the $\beta$-function
of the Non-Abelian Thirring model to first order in
$1/k$ in the large-$k$ expansion,
higher order terms in this  large-$k$ expansion.
These will not in general be absent, as our work presented in
this paper shows. Work along these lines is in 
progress.

\appendix \section*{Appendices}
\addtocontents{toc}{\protect\contentsline{section}
{\protect\numberline{ }\protect\large \protect\bf Appendices}{ }}

\section{Factorization of chain-diagrams}\label{factor chains} In this
appendix, we show how chain-diagrams factor, restricting ourselves to
2-loop order. This is done in appendix \ref{app:2loopfac}.  As a tool,
we need the ``conformal mapping technique'', which was introduced in
\cite{WieseDavid1995,WieseDavid1997}, reviewed in \cite{WieseHabil},
and which we present here for completeness, and since in contrast to
the cited references we here work exactly at the critical dimension,
where we need both an ultraviolet and an infrared cutoff.

\subsection{Factorization of chain-diagrams at 2-loop order}
\label{app:2loopfac} At 2-loop order, everything can with the help of
the magic relation be reduced to the bubble-chain. The subtracted
2-loop diagram, i.e.\ the 2-loop diagram minus the square of the
1-loop diagram is (we denote by $\cal S$ this subtraction-operator,
which also contains the integration and the cut-off functions)
\begin{equation}\label{lf26}
{\cal S}\left[\diagram{2banA}  \right]  = \int _{z,w}
\frac{1}{|z|^{2}|w|^{2}} \left[\Theta 
(a<|z|,|w|,|z-w|<L) -\Theta (a<|z|,|w|<L ) \right] \ .
\end{equation}
The first term on the r.h.s.\ represents the 2-loop integral, the
second term the subtracted 1-loop integrals (where integration over
$w$ and $z$ factorizes). 
Applying $-a\frac{\partial }{\partial a}$ to the above gives 
\begin{eqnarray}\label{lf4}
&&\!\!\!-   a\frac{\partial }{\partial a}\, {\cal
S}\!\left[\diagram{2banA}  \right] \nn \\
&& =  a \int _{z,w} 
\frac{1}{|z|^{2}|w|^{2}}\times\Big[\Theta (a=|z|<|w|,|z-w|<L)\nn \\
&&\hphantom{ a \int _{z,w} \frac{1}{|z|^{2}|w|^{2}} \times  \Big[\ }+ \Theta
(a=|w|<|z|,|z-w|<L)+ \Theta (a=|z-w|<|w|,|z|<L) \nn \\
&&\hphantom{ a \int _{z,w} \frac{1}{|z|^{2}|w|^{2}} \times  \Big[\ }-\Theta
(a=|z|<|w|<L ) -\Theta
(a=|w|<|z|<L ) \Big]\label{lf27}
\ .
\end{eqnarray}
Using the conformal mapping technique of
\cite{WieseDavid1995,WieseDavid1997,WieseHabil}, which is summarized
in appendix \ref{confmap}, all terms can be mapped onto $|z|=a$; with
the result (we have used that $a^{2}/|z|^{2}=1$):
\begin{eqnarray}
\!\!&-a&\frac{\partial }{\partial a}  ~ {\cal
S}\!\left[\diagram{2banA}  \right]\nn \\
\!\!&&\quad  =
\int_{w} 
\frac{1}{|w|^{2}}\left[
\Theta \left( \maxmin (|z|=a,|w|,|w-z|)<{\textstyle \frac{L}{a}}
\right) - \Theta 
\left(  \maxmin (a,|w|)<{\textstyle  \frac{L}{a}} \right)  \right]
\ .\qquad \label{lf28}
\end{eqnarray}
The function $\maxmin (a_{1},\dotsb,a_{n})$ is defined as
\begin{equation}\label{maxmin}
\maxmin (a_{1},\dotsb,a_{n}) := \frac{\mbox{max}
(a_{1},\dotsb,a_{n})}{\mbox{min} (a_{1},\dotsb,a_{n})}\ .
\end{equation}
The above is a function of $L/a$, and can be bounded for $L/a>c_{1}$ by
\begin{equation}\label{lf29}
\left|-a\frac{\partial }{\partial a}  ~ {\cal
S}\!\left[\diagram{2banA}  \right]  \right| <  c_{2} \frac{a}{L }
\end{equation}
with $c_{2}=c_{2} (c_{1})$. Taking $c_{1}>3$ allows the bound
$c_{2}=2$. The important thing is that the integral is not diverging:
this means we have subtracted the right 1-loop
counter-term. Moreover, the limit of large $L/a$ can be taken; since
it is zero, there is no single $\ln$-contribution in the 2-loop
integral. We can denote symbolically the result as
\begin{equation}\label{lf30}
\int\!\!\!\int \Diagram{2banA} ~=  \left[\int\Diagram{1ban}~ \right]^{2}\ . 
\end{equation}

\subsection{``Conformal mapping''} \label{confmap} 
As a tool to prove factorization of chains (see appendix
\ref{app:2loopfac}), we need the ``conformal mapping'' technique,
which was introduced in \cite{WieseDavid1995,WieseDavid1997}, reviewed
in \cite{WieseHabil}, and which we present here for completeness, and
since in contrast to the cited references we here work exactly at the
critical dimension, where we need both an ultraviolet and an infrared
cutoff.

Note that a general $N$-loop integral ${\cal I}_{N}$ will behave as 
\begin{equation}\label{lf31}
{\cal I}_{N} (a,L) = a_0 + a_{1}\ln\textstyle  \frac{L}{a}  + a_{2} \left(\ln
\frac{L}{a} \right)^{2} + \dots + a_{N}\left( \ln
\frac{L}{a}   \right)^{N}
\ , 
\end{equation}
where we dropped terms which vanish in the limit of $L/a\to \infty $. 
Deriving w.r.t.~$a$ leads to 
\begin{equation}\label{lf32}
-a \frac{ \partial }{\partial a} {\cal I}_{n} (a,L) = a_{1}+ 2 a_{2}
\ln\textstyle  \frac{L}{a} + \dots + N  a_{N}\left( \ln
\frac{L}{a}   \right)^{N-1}
\ .
\end{equation}
On the level of the integral, this operation amounts to fixing the
smallest distance to be $a$. Due to our normalizations, this is
equivalent to fixing the both endpoints of this smallest
distance. The integration over the remaining points has  then to be done. 

We now state a very important theorem for the integral over a
function $f$ at order $N-1$ loops: 
If $f (z_{1}, \bar z_{1}, \dots, z_{N},\bar z_{N})$ is a homogeneous function of
dimension  $-2 (N-1)$ ($z$ and $\bar z$ have dimension 1), then the
integral  over $z_{1}, \dots , z_{N-1}$ (the relative coordinates
between points) 
\begin{equation}\label{lf33}
\mathcal I_{N} (a,L) := \int_{z_{1},\dots, z_{N-1}} f (z_{1}, \bar z_{1}, \dots
,z_{N},\bar z_{N} ) {\cal C} (z_{1},\bar z_{1}, \dots,
z_{N}, \bar z_{N}) 
\end{equation}
has dimension 0. Consider a  sector $\cal S$ (ordering of the
distances).  Be $ x_{\alpha }:=|z_{i}-z_{j}|$, with $1\le \alpha \le m
:= N (N-1)/2$.  Then ${\cal S}:=\{z_{1}, \dots, \bar z_{N}\},\
\mbox{s.t.}\ x_{1}<x_{2}< \dots < x_{m}$. (Actually, we have chosen
the labeling of the distances $x_{\alpha }$ to account for the
ordering. This is not always the most practical thing to do.)  Also
define the characteristic function $\chi _{\cal S} (x_{1},\dots , 
x_{m})$ of a sector $\cal S$ as being 1 if all distances satisfy the
inequalities of the sector and 0 otherwise. The $a$-derivative of the
integral restricted to the sector $\cal S $ is
\begin{equation}\label{lf34}
{\cal J}^{\cal S}:= -a \frac{\partial }{\partial a} {\cal I}_{N}^{\cal
S} (a,L) = \int f (z_{1}, \dots, \bar z_{N} )\ts _{x_{1}=a} \Theta
(x_{m}<L) \chi _{\cal S} (x_{1}, \dots , x_{m})\ .
\end{equation}
The conformal mapping theorem
\cite{WieseDavid1995,WieseDavid1997,WieseHabil}, whose proof we
reproduce below for completeness, now states that {\em if} the
integral (\ref{lf34}) is Riemann-integrable everywhere, then
\begin{equation} \label{mapped}
{\cal J}^{\cal S} \equiv \int f
(z_{1}, \dots, \bar z_{N} )\ts _{x_i=a} \Theta (x_{m}/x_{1}<L/a) \chi
_{S} (x_{1}, \dots ,  x_{m})\ .
\end{equation}
In words: The above integral can be evaluated by fixing any of the
distances to be $a$ (or 1 equivalently). The constraint on the
smallest and largest  distances is captured by the condition that the
ratio of  largest to smallest  distance is bounded by $L/a$, as it is
in the original integral, which is thus just a special case of the
expression (\ref{mapped}).
\medskip 

\leftline{\underline{Proof:}}\noindent First of all, since $x_{1}=a$,
and introducing a $\delta$-function to enforce it, ${\cal J}^{\cal S}$
becomes
\begin{equation}\label{p1}
{\cal J}^{\cal S} = \int f (z_{1}, \dots,
\bar z_{N}  ) \delta (x_{1}- a) \Theta (x_{m}/x_{1}<L/a)  \chi
_{\cal S} 
(x_{1}, \dots ,  x_{m}) \ .
\end{equation}
We now aim at integrating over distances $x_{1},\dots , x_{m}$ instead of
coordinates with an arbitrary function $g$
\begin{equation}\label{p2}
\int \rmd^{2}  z_1 \dots \rmd^{2} z_{N-1}\, g (x_{1},\dots , x_{m}) =\int 
\rmd x_1 \dots  \rmd x_m \ 
  \mu (x_{1},\dots , x_{m}) g (x_{1},\dots , x_{m})\ .
\end{equation}
The measure is easily constructed as
\begin{equation}
 \mu (x_{1},\dots , x_{m}) =\int \rmd ^2
z_{1} \dots \rmd^{2} z_{N-1}\, \delta (x_{1}-|{z_{1}-z_{2}}|)\dots
\delta (x_{m}- |z_{N-1}-z_{N}|) \ ,\ 
\end{equation}
where the $\delta$-distributions enforce the $x_{i}$'s to be the
distances between the $z_{j}$'s.

We now want to map onto $x_{l}=a$. To achieve this, we can always do
the integration over $x_l$ last. This gives for ${\cal J}^{\cal S}$
\begin{eqnarray}\label{p3}
{\cal J}^{\cal S} =  \int \rmd x_{l}
&\displaystyle \int& \rmd x_1 \dots \rmd x_{l-1} \rmd x_{l+1} \dots
\rmd x_m\
 \mu  (x_{1},\dots , x_{m})\, \delta (x_{1}-a )\nonumber \\
&& \times \,f (x_{1},\dots , x_{m}) \,\Theta (x_{m}/x_{1}<L/a)\, \chi
_{\cal S} (x_{1}, \dots ,  x_{m})\ .\qquad \qquad
\end{eqnarray}
We now make a change of variables. For all $i$ but $l$, set
\begin{equation}\label{p4}
x_{i}:= \tilde x_{i} x_{l}/a\ . 
\end{equation}
We also define $\tilde x_{l}:=a $, and introduce this into
(\ref{p3}) as $1=\int \rmd \tilde x_{l}\, \delta
(\tilde x_{l} -a)$: 
\begin{eqnarray}\label{p5}
{\cal J}^{\cal S} = \int \rmd x_{l}
&\displaystyle \int& \rmd \tilde x_1 \dots 
\rmd \tilde  x_m\
 \mu  (\tilde x_{1}, \dots  ,\tilde  x_{m}) \, \delta
(\tilde x_{l} -a)\nonumber \\ 
&& \times \,f (\tilde x_{1}, \dots , \tilde x_{m})\, \Theta (\tilde
x_{m}/\tilde x_{1}<L/a)\, \chi _{\cal S}
(\tilde x_{1}, \dots , \tilde x_{m})\qquad \qquad\nonumber \\
&& \times\, \delta (\tilde x_{1}  x_{l}-a) \, \frac{a}{x_{l}}\ .
\end{eqnarray}
Note that the factor of $\frac a{x_{l}}$ consists of
$\left(\frac{x_{l}}{a} \right) ^{N (N-1)/2-1}$ 
from the terms $\rmd \tilde x_{i}$ but $\rmd \tilde x_{l}$; a factor
of $\left(\frac{x_{l}}{a} \right) ^{(N-1) (2-\frac{N}{2})}$ from the
measure; and a factor of $\left(\frac{x_{l}}{a} \right) ^{-2 ( N-1)}$
from $f$.  Using that
\begin{equation}\label{p6}
\int \rmd x_{l} \,  \delta (  \tilde x_{1}  x_{l} -a )
\frac{a}{x_{l}}  = 1\ ,
\end{equation} 
we obtain
\begin{eqnarray}\label{p7}
{\cal J}^{\cal S} = 
&\displaystyle \int& \rmd \tilde x_1 \dots 
\rmd \tilde  x_m \
 \mu  (\tilde x_{1}, \dots  ,\tilde  x_{m} ) \, \delta
(\tilde x_{l} -a )\nonumber \\ 
&& \times \,f (\tilde x_{1}, \dots , \tilde x_{m}) \Theta (\tilde  x_{m}/\tilde x_{1}<L/a)
\chi _{\cal S}
(\tilde x_{1}, \dots , \tilde x_{m})\ .\qquad \qquad 
\end{eqnarray}
Dropping the tildes, this is nothing but (\ref{p3}) with $x_1$
replaced by $x_{l}$ which completes the proof.


\section{The 4-loop integral}\label{4loopintegral}
In this Appendix we  evaluate analytically
 the integral (\ref{y5})
needed in Section (\ref{4loop simplified})
to obtain the {\em universal} part of the
4-loop contribution to the $\beta$-function [i.e.\ the last term
in (\ref{beta-4-loop-unspec})], with the result quoted
in (\ref{result 4loop integral}).
The integrand  of the integral (\ref{y5}) in question is
\begin{equation}\label{lf35}
{\cal M}:=
\left(\frac{1}{z_{14}\bar z_{13}}- \frac{1}{\bar z_{14} z_{13}}\right)
\left(\frac{1}{z_{24}\bar z_{23}}- \frac{1}{\bar z_{24} z_{23}}\right)
\frac{1}{|z_{34}|^{2}} \left(\frac{1}{z_{15}\bar z_{25}}+
\frac{1}{\bar z_{15} z_{25}} \right) \ .
\end{equation}
Graphically, this is depicted in figure \ref{f:4lMOPE}.  We observe
that we can make the following simplification (due to the ``second
magic rule''):
\begin{equation}\label{lf36}
\frac{1}{w\bar u} - \frac{1}{\bar w u} = \frac{\bar w u -w\bar u}{\bar
ww \bar uu}=  2i\frac{\vec{w}\times \vec{u}  }{
|w|^{2} |u|^{2}} =  2i \frac{ |\vec{w}-\vec{u}| h   }{|w|^{2} |u|^{2}}\ ,
\end{equation}
where $h$ is the height of the triangle spanned by $\vec{w} $ and
$\vec{u} $; if the angle is larger than $\pi $, then $h$ is negative.
Graphically, this can be visualized as
\begin{equation}\label{lf37}
\vardiagram{.27}{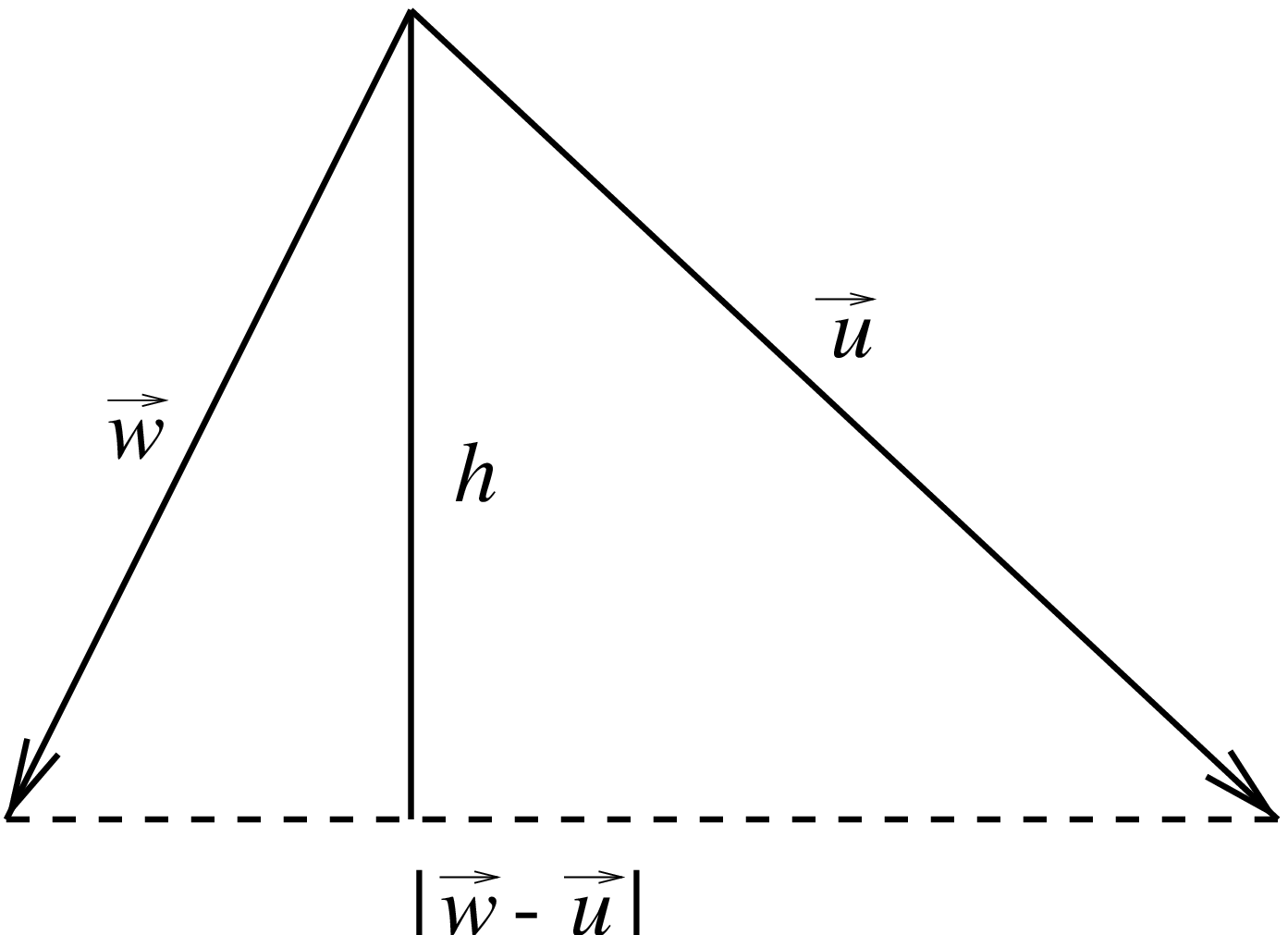}
\end{equation}
Note that the first two factors of the integrand $\cal M$ both
contribute a term $|z_{34}|$, thus canceling the third term
$\frac{1}{|z_{34}|^{2}}$.  This allows us to see that the integral has no
subdivergences; it will contain only a ``global divergence'',
i.e.\ it will be proportional to a single power of $\ln(L/a)$ ($L$ and
$a$ are the IR and UV cutoffs, respectively).  We now proceed to check
this by  explicit calculation and to compute the precise coefficient
of the single logarithmic divergence.  Let us now introduce distances
as depicted in the figure:
\begin{equation}\label{lf38}
\vardiagram{.27}{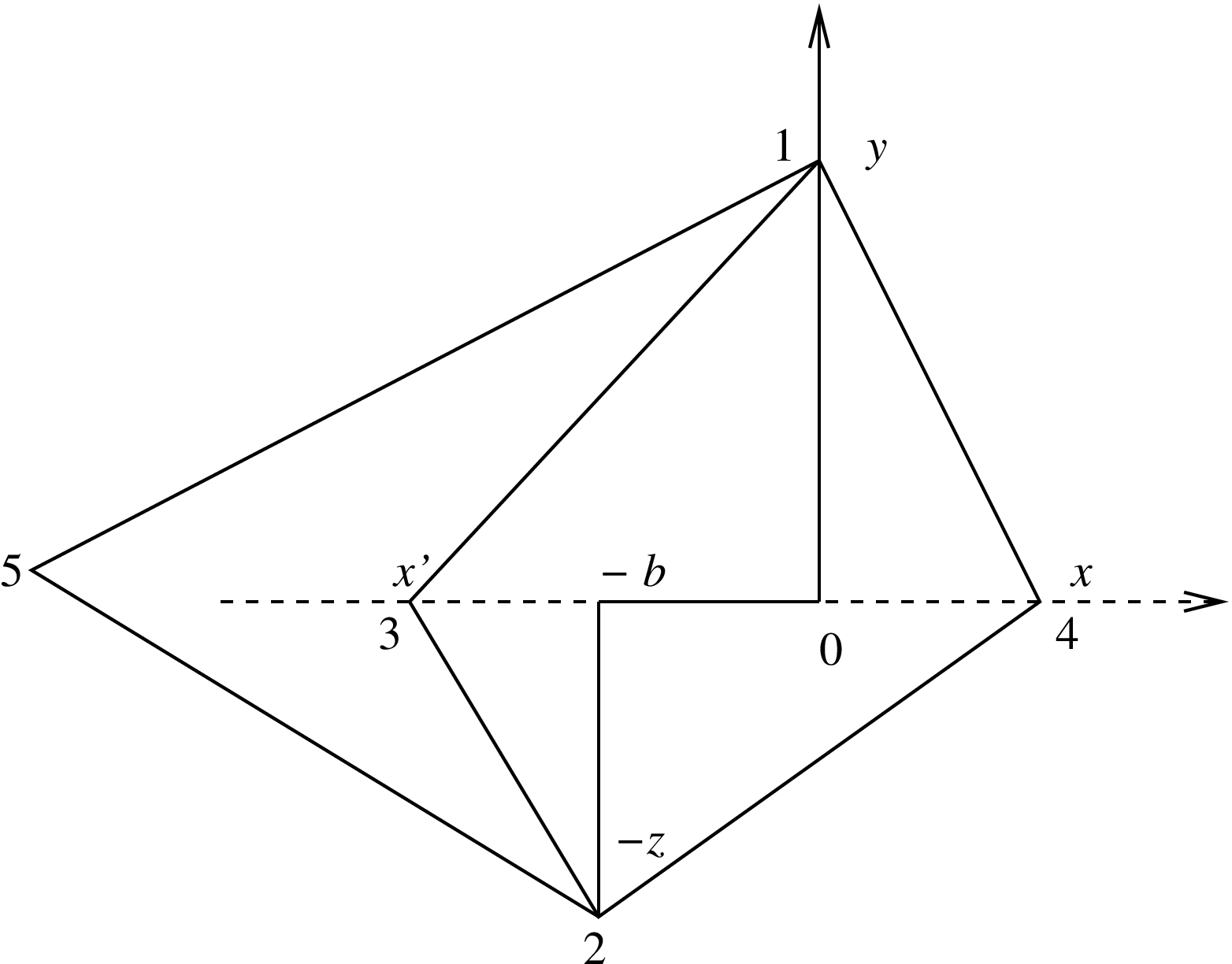}
\end{equation}
Here all distances are measured from $0$ except for $x'$ and $z$ which
are measured from their intersection point.  In these conventions,
$x'$ and $-z$ in the figure are negative.  The integrand can then be
written as \footnote{There are four complex integration variables,
equivalent to eight real integration variables. We make use of this
equivalence whenever convenient.}
\begin{equation}\label{Msimp}
{\cal M}=
(2i)^{2} \frac{y}{(y^{2}+x^{2}) (y^{2}+ (x'+b)^{2})}
\frac{- z}{(z^{2}+{ x'}^{2}) (z^{2}+ (x+b)^{2})}
 \left(\frac{1}{z_{15}\bar z_{25}}+
\frac{1}{\bar z_{15} z_{25}} \right)\ ,
\end{equation}
where all variables are to be integrated over. Let us first do the
integrals over $x$, $ x'$, $b$, $z$ and $ z_5=\Re(z_5) + i \Im(z_5)$,
i.e.\ all distances except for $y$, which is kept {\em fixed and
positive}.  At the end, we integrate over the vector $y$, both over
{\em its magnitude and direction}. (This fixes the coordinate system.)  We
note that choosing $a\ll |y| \ll L$, boundary terms can be neglected,
since the integrals do not contain subdivergences, neither in the UV
nor in the IR.

Doing first the integral over point 5, we obtain
using (\ref{A.1}) from appendix \ref{app:Integrals}
\begin{equation}\label{lf39}
\int \rmd ^2 z_{5}\,  \left(\frac{1}{z_{15}\bar z_{25}}+
\frac{1}{\bar z_{15} z_{25}} \right) = -2 \pi \ln (|z_{12}|^{2})+\mbox{const.} = -2\pi \ln \left[(y+z)^{2}+b^{2} \right]+\mbox{const.}\ ,
\end{equation}
where the constant depends on the IR-cutoff $L$. However one easily sees
that it drops from the above calculation due to the asymmetry of $z\to
-z$ of the remaining terms in (\ref{Msimp}). 
We finally have to integrate:
\begin{equation} \label{M2}
-8 \pi\, \frac{y}{(y^{2}+x^{2}) (y^{2}+ (x'+b)^{2})}
\frac{z}{(z^{2}+{ x'}^{2}) (z^{2}+ (x+b)^{2})}
 \ln\left[(y+z)^{2}+b^{2} \right]\ .
\end{equation}
The simplest integrals are those over $x$ and $x'$. We use
\begin{equation}\label{lf40}
\int_{-\infty }^{\infty }\rmd x\, \frac{1}{x^{2}+y^{2}}
\frac{1}{(b+x)^{2}+z^{2}} = \frac{\pi (|y|+|z|) }{|yz| (b^{2}+
(|y|+|z|)^{2})} \ ,
\end{equation}
which can be done by residue-calculus.
Integrating (\ref{M2}) over $x$ and $x'$ thus gives:
\begin{equation}\label{lf41}
-8 \pi^{3} \,
 \frac{ (|y|+|z|)^{2} }{y z (b^{2}+ (|y|+|z|)^{2})^{2}}
 \ln\left[(y+z)^{2}+b^{2} \right]\ .
\end{equation}
To continue, we recall that by {\em construction} $y$ (which is the
module of a vector) is
positive. (\ref{lf41}) can thus be written as the integral over
positive $z$ only
\begin{equation}\label{lf42}
-8 \pi^{3} \,
 \frac{ (y+z)^{2} }{y z (b^{2}+ (y+z)^{2})^{2}}
\left(  \ln\left[(y+z)^{2}+b^{2} \right] - \ln\left[(y-z)^{2}+b^{2}
\right] \right) 
\end{equation}
The easiest integral to do is that over $b$, which nevertheless is a
little bit tricky. We need
\begin{equation}\label{lf43}
\int_{-\infty }^{\infty } \frac{\ln (b^{2}+d^{2})}{(b^{2}+s^{2})^{2}}\,
\rmd b =  \frac{-\pi }{s^{2} (|d|+|s|)} + \frac{\pi \ln (|d|+|s|)}{|s|^{3}}\ .
\end{equation}
which can be verified with the help of the residue-theorem. To do so,
one splits the $\ln (b^{2}+d^{2}) = \ln (b+i |d|)+\ln (b-i |d|)$ which
both have branch-cuts. But the integral can be closed either in the
upper or lower domain, and we close it in the domain where there is no
branch-cut.
This leaves us with 
\begin{eqnarray}\label{M3}
8 \pi ^{4}\int_{0}^{\infty} \rmd z\, \frac{(y+z)^{2}}{yz}&\bigg\{&
\left[\frac{1 }{(y+z)^{2} 
(2 |y+z|)} - \frac{ \ln (2 |y+z|)}{(y+z)^{3}} \right]\nn \\
&&
-\left[\frac{1 }{(y+z)^{2} 
(|y-z|+|y+z|)} - \frac{ \ln (|y-z|+|y+z|)}{|y+z|^{3}} \right] \bigg\}\ .\nn \\
\end{eqnarray}
Scaling out $y$, and splitting the integral into domains where the
absolute values have a definite sign gives:
\begin{eqnarray}\label{M4}
\frac{8 \pi ^{4}}{y^{2}}\int_{0}^{\infty} \rmd z\,
\frac{1}{z}&\bigg\{& 
\left[\frac{1 }{ 
2 |1+z| } - \frac{ \ln (2 |1+z|)}{(1+z)} \right]\nn \\
&&
-\left[\frac{1 }{ 
(|1-z|+|1+z|)} - \frac{ \ln (|1-z|+|1+z|)}{|1+z|} \right]
\bigg\}\nn \\
=\frac{8 \pi ^{4}}{y^{2}}\int_{0}^{1} \rmd z\,
&\bigg\{& 
\left[\frac{1 }{2z (1+z) 
} - \frac{ \ln (2 ( 1+z))}{z(1+z)} \right]
-\left[\frac{1 }{2 z } - \frac{ \ln (2)}{z( 1+z)} \right]
\bigg\}\nn\\ 
+\frac{8 \pi ^{4}}{y^{2}}\int_{1}^{\infty} \rmd z\,
&\bigg\{& 
\left[\frac{1 }{2z (1+z) 
} - \frac{ \ln (2 ( 1+z))}{z(1+z)} \right]
-\left[\frac{1 }{2z^{2} } - \frac{ \ln (2 z)}{z( 1+z)}
\right] \bigg\}\nn \\
 &&\hspace{-2.5cm}=- \frac{2}{3}\pi ^{4} \left(6+\pi ^{2} \right)
\frac{1}{y^{2}} 
\end{eqnarray}  
The final integral over $y$ contains the integral over the modulus of
$y$ and its direction, which contributes a factor of $2\pi$:
\begin{equation}\label{lf44}
\int_{a}^{L}\rmd y\, 2\pi y\,\left[ - \frac{2}{3}\pi ^{4} \left(6+\pi ^{2}
\right) \frac{1}{y^{2}}\right] =
-\frac{4}{3}\pi ^{5}  \left(6+\pi ^{2} \right)
\, \ln\!\left(\frac{L}{a} \right) \ .
\end{equation}
To conform to the  normalizations used in the main text, see Eq.\
(\ref{actionNATM}), this still has to be divided by $(2\pi )^{4}$,
yielding the final result [with the integral running over all but one
of the points, and normalizations according to equation
(\ref{actionNATM})]
\begin{equation}\label{K1}
{\cal I}:= \int {\cal M} = -\frac{\pi }{12}  \left(6+\pi ^{2} \right)
\, \ln\!\left(\frac{L}{a} \right) +\mbox{finite}\ .
\end{equation}
We have indicated an additional finite term in the result, which
depends on the specific regularization prescription, and which is
either a constant or decays to 0 in the limit of $L/a\to \infty$.

\section{Some remarks on group theory}\label{remarksgrouptheory}
In this appendix, we collect a number of useful group-theoretical
identities, first in appendix \ref{a:algebra} for a general Lie-group $\cal G$,
then in appendix \ref{SUn} for $\mbox{SU} (N)$,  and finally in
appendix \ref{otheralgebras} for  the other classical groups,
$\mbox{SO}(N)$ and $\mbox{SP}(N)$.

\subsection{Group theoretical invariants}\label{a:algebra} In this
appendix we discuss the additional group theoretical invariant,
referred to in the main text. Since we are using the current algebra,
only the adjoint representation of the symmetry group ${\cal G}$
appears in our calculations.  Therefore, all group-theoretical
invariants that can possibly appear, can all be constructed out the
structure constants. The simplest such invariant is of course the
eigenvalue of the quadratic Casimir invariant $C_2$ in the adjoint
representation, which is of second order in structure constants
$f^{abc}$.  Here we consider invariants which are of higher order in
the structure constants.

\medskip

{\it Notation}: \quad The zero modes $j^a := J_0^a$ $= \oint
(dz/2\pi i) J^a(z)$ of the Kac-Moody currents
\cite{KnizhnikZamolodchikov1984} are the generators of the Lie-group
${\cal G}$, satisfying the commutation relations
\begin{equation}\label{lf45}
\left[j^{a},j^{b} \right] = {f_{c}}^{ab} j^{c} \ ,
\end{equation}
which are represented in
the adjoint representation by matrices
\begin{equation}
\label{AdjointReprMatricesT}
{(T^a)_c}^b := {f_{c}}^{ab}
\end{equation}
We work with {\it antihermitean} generators  $j^a$, so that  the
structure constants $ {f_{c}}^{ab}$ are  real.

The  $\cal G$-invariant Killing form
$\eta^{ab}$, and its inverse $\eta_{bc}$, defined by
\begin{equation}
\label{Killing}
\eta^{ab}:={-1\over {\cal N}  }\, \tr \left ( T^a {T^b}\right ) \ ,
\qquad \eta^{ab} \eta_{bc} = \delta^a_c
\end{equation}
may be used to raise and lower adjoint indices $a,b,\dots$.  Here
${\cal N}$ is a suitable normalization constant.  Then, (\ref{lf45})
and (\ref{Killing}) imply that the structure constants $ f^{cab}=
f^{abc} $ are totally antisymmetric.  Throughout this subsection, we
choose a basis of the Lie algebra for which $\eta^{ab}=\delta^{ab}$.
Hence, no distinction between upper and lower adjoint indices has to
be made.  [The matrices ${(T^a)_c}^b$ in (\ref{AdjointReprMatricesT})
are then antihermitean.]

We now proceed to discuss various group-theoretical invariants, needed
in the main text, which can be constructed out of products of
structure constants.  Our discussion is organized according to the
number of factors $f^{abc} $ appearing.

\medskip 

{\it Quadratic Casimir}: \quad The eigenvalue $C_2$ of the
 quadratic Casimir invariant in the adjoint
representation\footnote{As usual, all repeated indices are summed.}, 
\begin{equation}
\label{QadraticCasimirDef} {{\left ( T^a T^a \right )}_c}^d =
{f_{c}}^{ab} \ {f_{b}}^{ad} = - C_2 \, {\delta_c}^d \qquad
\Leftrightarrow \quad f^{abc} \ f^{abd} = C_2 \ \delta^{cd}
\end{equation}
is of 2nd order in the structure constants.  Eq.\
(\ref{QadraticCasimirDef}) is graphically depicted in (\ref{alg1}).

\medskip 

{\it ``Triangle Rule''}: The Jacobi-identity
implies the following relation for the structure constants:
\begin{equation} \label{Jacobi2}
{f_{e}}^{ad} {f_{d}}^{bc} + {f_{e}}^{bd} {f_{d}}^{ca} + {f_{e}}^{cd} {f_{d}}^{ab}
 = 0
\ .
\end{equation}
which is just (\ref{lf45}):
\begin{equation}
\label{lf51}
{\left ( [T^a,  T^b] \right )_e}^c
 = {f_{d}}^{ab} \ {\left (T^d \right )_e}^c
\end{equation}
Multiplying (\ref{Jacobi2}) with ${f^{g}}_{ab}$, yields
\begin{equation}\label{lf48}
0=
{f^{g}}_{ab}
 \left[
f^{ead} {f_{d}}^{bc} + f^{ebd} {f_{d}}^{ca} + f^{ecd} {f_{d}}^{ab}
 \right]
= - \tr \left ( T^g T^{c}T^{e} \right) -  \tr \left ( T^{c}T^{e}T^g
\right) + C_{2} f^{ecg}
 \label{lf49}
\ .
\end{equation}
Using the cyclic invariance of the trace, this yields  the ``triangle rule''
\begin{equation}\label{Wigner}
 \tr \left ( T^{g}T^{c}T^{e} \right) ={-1\over 2} C_2
f^{gce}
\ .
\end{equation}
Eq.\ (\ref{Wigner}) is graphically depicted in
(\ref{alg2}).
\medskip

{\it Invariant 4-index tensor $d^{abcd}$}: \quad Next we
consider the following totally symmetrized trace of four (adjoint)
representation matrices
\begin{equation}\label{dabcd}
 d^{abcd}:=   \tr  \left(T^{\{a}T^{b}T^{c}T^{d \}} \right)\ ,
\end{equation}
which is ${\cal G}$-invariant by construction.  This invariant arises
when considering traces of four matrices $T$,
Eq.~(\ref{AdjointReprMatricesT}). The result is given in (\ref{4Tred})
below. To derive it, observe that for traces of more than three
generators $T$, which cannot be reduced using (\ref{Wigner}), one can
permute two $T$'s, with the aim of creating a loop of 3 with the
remaining $T$'s, which, in turn, can then be reduced using
(\ref{Wigner}).  For a trace of four $T$'s, this reads
\begin{equation}\label{4Ts}
\displaystyle \tr  \left(T^{a}T^{b}T^{c}T^{d} \right) = -\half
C_2 f^{abh}f^{cdh} + \tr \left(T^{b}T^{a}T^{c}T^{d} \right)\ .
\end{equation}
This also tells us that
\begin{equation} \label{4T2}
\tr  \left(T^{a}T^{b}T^{c}T^{d} \right) = \tr
\left(T^{b}T^{a}T^{d}T^{c} \right)=\tr \left(T^{d}T^{c}T^{b}T^{a} \right)
\end{equation}
where the second relation is obtained using the cyclic invariance of
the trace.  We now want to calculate a general trace of four
$T$'s. First, by using (\ref{4T2}), and the cyclic invariance of the
trace, we find that of the 6 possible permutations, which leave the
first index unchanged only 3 are independent. These are $ K_1 = \tr
\left(T^{a}T^{b}T^{c}T^{d} \right) =\tr \left(T^{a}T^{d}T^{c}T^{b}
\right) $, $ K_2 = \tr \left(T^{a}T^{c}T^{d}T^{b} \right) =\tr
\left(T^{a}T^{b}T^{d}T^{c} \right) $, and $ K_3 = \tr
\left(T^{a}T^{d}T^{b}T^{c} \right) =\tr \left(T^{a}T^{c}T^{b}T^{d}
\right)$.  The totally symmetrized trace, defined in (\ref{dabcd}),
can now be expressed in terms of the $K_{i}$ as: $d^{abcd}=\frac{1}{3}
(K_{1}+K_{2}+K_{3})$. Writing $K_{1}=d^{abcd}+\frac{1}{3}\left[(
K_{1}-K_{2})+ (K_{1}-K_{3}) \right] $, and using (\ref{lf51}), we can
rewrite each of these terms with the help of $d^{abcd}$ and $f$'s as
\begin{equation}\label{4Tred}
  \tr \left(T^{a}T^{b}T^{c}T^{d} \right) =
d^{abcd}+\frac{C_{2}}{6}\left[f^{adh}f^{bch}-f^{abh} f^{cdh} \right]
\end{equation}
The invariant $d_2$ is now defined by
\begin{equation}\label{lf57}
\frac{1}{{\cal N}_{\ind{ad}}} d^{abcd}d^{abcd}
=\frac{C_{2}^{4}}{24}+d_{2}
\end{equation}
where ${\cal N}_{\ind{ad}}$ is the dimension of the adjoint representation.
Note that   (\ref{4Tred}), (\ref{lf57}), (\ref{Wigner}) imply
\begin{equation}
\label{cube-invariant} \tr \left(T^{a}T^{b}T^{c}T^{d} \right) \, \tr
\left(T^{a}T^{b}T^{c}T^{d} \right) = {\cal N}_{\mathrm{ad}} \left ( {
C_2^4\over 8} + d_2 \right )
\end{equation}
The l.h.s.\ can graphically be viewed as the ``cube''-invariant,
discussed in section \ref{present results}, and depicted in
Fig.~\ref{cube} of the same section [recall
(\ref{AdjointReprMatricesT})].

In  section \ref{SUn}, we  show that for  $\mbox{SU} (N)$
\begin{equation}\label{lf58}
d_{2}= \frac{3}{2}\,N^{2}\ .
\end{equation}
(The quadratic Casimir is $C_{2}=N$ in our conventions.) This is
in agreement with the results of Ref.\cite{AliGracey2001}.  Hence, for
$\mbox{SU}(N)$, the term $d_2$ in (\ref{lf57}) is subleading in $N$,
as compared to the first term.  This subleading $N$-dependence of
$d_2$ is also true for all the remaining classical groups, which
follows from (\ref{lf57}) and (\ref{C2d2OtherGroups}).

\subsection{$\mbox{SU} (N)$}\label{SUn} In this section we present a
derivation of the value of the invariant $d_2$ for ${\cal
G}=\mbox{SU}(N)$, i.e.\ of (\ref{lf58}), which provides an independent
check of this result given in Ref.\cite{AliGracey2001}\footnote{This
method can also be used to calculate higher invariants
\cite{LudwigWieseUnpublished}.}.

We start by recalling the generators in the (complexified) Lie algebra
of $\mbox{SU}(N)$ in the {\it fundamental} representation
\begin{equation}\label{lf59}
X^{a} \equiv  X^{\alpha }_{\bar \alpha } :=
\{ \mbox{matrix with 1 in column
$\alpha $, row $\bar \alpha $; 0 elsewhere} \}\ ,
\qquad (\alpha,{\bar \alpha} = 1, ..., N)
\end{equation}
where the adjoint index $a=\left\{{}^{\alpha }_{\bar\alpha}\right\}$.
These satisfy
\begin{equation}\label{lf61}
\left[X^{\alpha }_{\bar \alpha } , X^{\beta }_{\bar \beta } \right]
= \delta ^{\alpha}_{\bar \beta } X^{\beta }_{\bar \alpha} -
\delta^{\beta }_{\bar \alpha}  X^{\alpha}_{\bar \beta }\ ,
\end{equation}
which yields the structure constants in this basis
\begin{equation}\label{lf62}
{f_c}^{ab}\equiv f^{{\bar \gamma} \alpha \beta }_{\gamma \bar \alpha
\bar \beta } = \delta ^{\bar \gamma }_{\bar \alpha } \delta ^{\beta
}_{\gamma }\delta ^{\alpha }_{\bar \beta } - \delta ^{\alpha }_{\gamma
} \delta ^{\bar \gamma }_{\bar \beta } \delta ^{\beta }_{\bar \alpha }
\ .
\end{equation}
The Killing form is given by 
\begin{eqnarray}
\eta ^{ab} &\equiv & \eta ^{\alpha \beta }_{\bar \alpha\bar
\beta } := \frac{-1}{N}\tr 
\left(T^{a}T^{b}\right)=\frac{-1}{N} \left({f_{d}}^{ac}{f_{c}}^{bd}
\right) = - 2 \left(\delta ^{\alpha }_{\bar \beta }\delta ^{\beta
}_{\bar \alpha } -\frac{1}{N} \delta ^{\alpha }_{\bar \alpha }\delta
^{\beta }_{\bar
\beta }\right) \nn \\
&=&- 2 \left(\mbox{ projector onto the adjoint} \right)\label{lf64}
\end{eqnarray}
and its inverse
\begin{equation}\label{lf65}
\eta _{ab}
=
\eta
 _{\alpha \beta }^{\bar \alpha\bar \beta  }
=  \frac{-1}{2} \left(\delta _{\alpha }^{\bar \beta }\delta _{\beta }^{\bar \alpha }
-\frac{1}{N} \delta _{\alpha }^{\bar \alpha }\delta _{\beta }^{\bar
\beta }\right) \ .
\end{equation}
One easily finds 
\begin{equation}\label{lf67}
\eta _{ab}\eta^{ba}  = (N-1) (N+1) = {\cal
N}_{\mathrm{ad}} =\mbox{dimension of adjoint representation}\ .
\end{equation}
Since we use $\eta$ to raise and lower indices
$a=\left\{{}^{\alpha }_{\bar \alpha}\right\}$ of structure constants
which are traceless [see (\ref{lf62})], one can also use the
simplified form
\begin{equation}\label{lf66}
\eta _{ab}
~\to~ 
\eta _{ab}^{\mathrm{simp}}
=   \frac{-1}{2} \delta _{\alpha }^{\bar \beta }\delta _{\beta }^{\bar \alpha }
\end{equation}
instead of $\eta_{ab}$, for calculational convenience.  Writing
${{\left( T^a\right)}_c}^b= {f_c}^{ab} $ we obtain\footnote{Again we
use a computer to do the algebra.}:
\begin{eqnarray}
\frac{1}{{\cal N}_{\mathrm{ad}}} \tr \left(T^{a}T^{a'} \right)\eta _{aa'} &=& - N\label{lf68}\\
\frac{1}{{\cal N}_{\mathrm{ad}}} \tr \left(T^{a}T^{b} \right)
 \tr \left(T^{a'}T^{b'} \right)\eta _{aa'}\eta _{bb'} &=& N^{2}\label{lf69}\\
\frac{1}{{\cal N}_{\mathrm{ad}}} \tr \left(T^{a}T^{b}T^{c} \right)
 \tr \left(T^{a'}T^{b'}T^{c'} \right)\eta _{aa'}\eta _{bb'}\eta _{cc'}
&=& \frac{1}{4}N^{3} \label{C3}\\
\frac{1}{{\cal N}_{\mathrm{ad}}} \tr \left(T^{a}T^{b}T^{c}T^{d} \right)
 \tr \left(T^{a'}T^{b'}T^{c'} T^{d'}\right)\eta _{aa'}\eta _{bb'}
\eta_{cc'}\eta _{dd'}
&=& \frac{1}{8}N^{4}+\frac{3}{2}N^{2}\label{C4}\qquad
\end{eqnarray}
Comparison of (\ref{C4}) with (\ref{cube-invariant}) yields
$d_2={3\over 2} N^2$, in agreement with (\ref{cube-invariant}), and
(\ref{lf58}).

\subsection{Other groups}\label{otheralgebras}
Besides $\mbox{SU} (N)$ we also consider 
 $\mbox{SO} (N)$ and $\mbox{SP(N)}$. The results of
Ref.  \cite{vanRitbergenVermaserenLarin1997} yield:
\begin{equation}
\label{C2d2OtherGroups}
\begin{array}{rll}
\mbox{SU} (N)\,: &\qquad  C_{2}= N\qquad  & d_{2} = \frac{3}{2}N^{2} \\
\mbox{SO} (N)\,: &\qquad  C_{2}= N-2 \qquad  & d_{2} = 
24-\frac{45}{2}N+6N^{2}-\frac{3}{8}N^{3}\rule{0mm}{5mm}\\
\mbox{SP} (N)\,: &\qquad  C_{2}= \frac{N+2}{2}\qquad  & d_{2} =
\frac{3}{2} +\frac{45}{32}N+\frac{3}{8}N^{2}+\frac{3}{128}N^{3}\rule{0mm}{5mm}
\end{array}
\end{equation}
We have already quoted these values for the group theoretical
invariants on figure \ref{grouptable}, but repeat them here for the
convenience of the reader.  Note that one can always normalize the
1-loop coefficient in the $\beta$-function for $g$ [the term $\propto
g^2$ in (\ref{beta-4-loop-unspec})] to $1/2$, by rescaling $g$ by a
constant.  This means that the normalization-invariant quantity which
enters at 4-loop order is $d_{2}/C_{2}^{4}$.  This allows us to
perform the following checks on (\ref{C2d2OtherGroups}), by using well
known isomorphism between the corresponding Lie algebras:
\begin{eqnarray}\label{oa1}
\frac{d_{2}}{C_{2}^{4}}\lts_{\mathrm{SU} (2)} &=&
\frac{d_{2}}{C_{2}^{4}}\lts_{\mathrm{SO} (3)} = \frac{3}{8} \\
\frac{d_{2}}{C_{2}^{4}}\lts_{\mathrm{SO} (5)} &=&
\frac{d_{2}}{C_{2}^{4}}\lts_{\mathrm{SP} (4)} =
\frac{13}{72}\label{oa2} \\
\frac{d_{2}}{C_{2}^{4}}\lts_{\mathrm{SU} (4)} &=&
\frac{d_{2}}{C_{2}^{4}}\lts_{\mathrm{SO} (6)} =
\frac{3}{32}\label{oa3}\\
\frac{d_{2}}{C_{2}^{4}}\lts_{\mathrm{SO} (-N)} &=&
\frac{d_{2}}{C_{2}^{4}}\lts_{\mathrm{SP} (N)} = \frac{3 [32+N
(14+N)]}{8 (2+N)^{3}}\label{oa4} \ .
\end{eqnarray}

\section{Some elementary Integrals}
\label{app:Integrals}
In this appendix, we consider some elementary integrals,
quoted in the main text.
Consider two points 
$z_{a}$ and $z_{b}$ in the complex plane, which are well
inside a circle of (large) radius $R$ centered at the origin.
It is  then elementary to establish the following result:
\begin{equation}\label{A.1}
\int_{|\fz |\le R} \rmd ^2\fz \, \frac{1}{(\fz -z_{a}) (\fz^{*}-z_{b}^{*})} = 
- \pi \ln |z_{b}-z_{a}|^{2} + \pi \ln R^{2}+ \pi \ln
\left[1-\frac{z_{a}z_{b}^{*}}{R^{2}}  \right]
\ .
\end{equation}
Furthermore, for $|\fz -z_{a}|\le a\le |z_{a}-z_{b}|$ 
\begin{equation}\label{A.2}
\int_{|\fz -z_{a}|\le a} \rmd ^2\fz \, \frac{1}{(\fz -z_{a}) (\fz
^{*}-z_{b}^{*})} =0 
\ .
\end{equation}
Finally, this implies upon taking the
limit of $R\to \infty $,

\begin{equation}\label{A.3}
\int_{|\fz -z_{a}|,|\fz -z_{b}| \ge a} \rmd ^2\fz \left[\frac{1}{(\fz
-z_{a}) (\fz^{*}-z_{b}^{*})}-\frac{1}{|\fz -z_{a}|^{2}} \right] =
-2\pi \ln \left|\frac{z_{a}-z_{b}}{a} \right|
\ .
\end{equation}
 as long as $|z_{a}-z_{b}|\ge 2a$
(up to terms of order $a^{2}$ which are neglected).


\setcounter{section}{17} 
\section{References}
\def\refname{} 

\begin{thebibliography}{10}

\bibitem{KnizhnikZamolodchikov1984}
V.G. Knizhnik and A.B. Zamolodchikov,
\newblock {\em Current algebra and {Wess-Zumino} model in two dimensions},
\newblock Nucl. Phys. {\bf B 247} (1984)   83--103.

\bibitem{DashenFrishman1973}
R.~Dashen and Y.~Frishman,
\newblock {\em Thirring model with {U(n)} symmetry: scale invariant only for
  fixed values of a coupling constant},
\newblock Phys. Lett. B {\bf 46} (1973)   439--42.

\bibitem{DashenFrishman1975}
R.~Dashen and Y.~Frishman,
\newblock {\em Four-fermion interactions and scale invariance},
\newblock Phys. Rev. D {\bf 11} (1975)   2781--801.

\bibitem{GrossNeveu1974}
D.J. Gross and A.~Neveu,
\newblock {\em Dynamical symmetry breaking in asymptotically free field
  theories},
\newblock Phys. Rev. D {\bf 10} (1974)   3235--53.

\bibitem{LudwigFisherShankarGrinstein1994}
A.W.W. Ludwig, M.P.A. Fisher, R.~Shankar  and G.~Grinstein,
\newblock {\em Integer quantum hall transition: an alternative approach and
  exact results},
\newblock Phys. Rev. B {\bf 50} (1994)   7526--52.

\bibitem{EfetovBook}
K.~Efetov,
\newblock {\em Supersymmetry and disorder and chaos},
\newblock Cambridge University Press, Cambridge, U.K., 1997.

\bibitem{LevineLibbyPruisken1983}
H.~Levine, S.B. Libby  and A.M.M. Pruisken,
\newblock {\em Electron delocalization by a magnetic field in two dimensions},
\newblock Phys. Rev. Lett. {\bf 51} (1983)   1915--18.

\bibitem{LevineLibbyPruisken1984a}
H.~Levine, S.B. Libby  and A.M.M. Pruisken,
\newblock {\em Theory of the quantized {Hall} effect. {I}},
\newblock Nucl. Phys. B {\bf 240} (1984)   30--48.

\bibitem{LevineLibbyPruisken1984b}
H.~Levine, S.B. Libby  and A.M.M. Pruisken,
\newblock {\em Theory of the quantized {Hall} effect. {II}},
\newblock Nucl. Phys. B {\bf 240} (1984)   49--70.

\bibitem{LevineLibbyPruisken1984c}
H.~Levine, S.B. Libby  and A.M.M. Pruisken,
\newblock {\em Theory of the quantized {Hall} effect. {III}},
\newblock Nucl. Phys. B {\bf 240} (1984)   71--90.

\bibitem{Khmelnitskii1983}
D.E. Khmel'nitskii,
\newblock {\em Quantization of hall conductivity},
\newblock Zhurnal Eksperimental'noi i Teoreticheskoi Fiziki {\bf 38} (1983)
  454--8,
\newblock JETP Lett. {\bf 38} (1983) 552-6.

\bibitem{Zirnbauer1996}
M.R. Zirnbauer,
\newblock {\em Riemannian symmetric superspaces and their origin in
  random-matrix theory},
\newblock J. Math. Phys. {\bf 37} (1996)   4986--5018.

\bibitem{AltlandZirnbauer1997}
A.~Altland and M.R. Zirnbauer,
\newblock {\em Nonstandard symmetry classes in mesoscopic
  normal-superconducting hybrid structures},
\newblock Phys. Rev. B {\bf 55} (1997)   1142--61.

\bibitem{SenthilFisherBalentsNayak1998}
T.~Senthil, M.P.A. Fisher, L.~Balents  and C.~Nayak,
\newblock {\em Quasiparticle transport and localization in high-{$T_c$}
  superconductors},
\newblock Phys. Rev. Lett. {\bf 81} (1998)   4704--7.

\bibitem{SenthilFisher2000}
T.~Senthil and M.P.A. Fisher,
\newblock {\em Quasiparticle localization in superconductors with spin-orbit
  scattering},
\newblock Phys. Rev. B {\bf 61} (2000)   9690--8.

\bibitem{MudryChamonWen1996}
C.~Mudry, C.~Chamon  and X.~Wen,
\newblock {\em Two-dimensional conformal field theory for disordered systems at
  criticality},
\newblock Nucl. Phys. B {\bf 466} (1996)   383--443.

\bibitem{ChamonMudryWen1996}
C.~Chamon, C.~Mudry  and X.~Wen,
\newblock {\em Instability of the disordered critical points of dirac
  fermions},
\newblock Phys. Rev. B {\bf 53} (1996)   R7638--41.

\bibitem{GuruswamyLeClairLudwig2000}
S.~Guruswamy, A.~LeClair  and A.W.W. Ludwig,
\newblock {\em {$\mbox{gl}(N|N)$} super-current algebras for disordered dirac
  fermions in two dimensions},
\newblock Nucl. Phys. {\bf B 583} (2000)   475--512.

\bibitem{GerganovLeClairMoriconi2001}
B.~Gerganov, A.~LeClair  and M.~Moriconi,
\newblock {\em Beta function for anisotropic current interactions in 2d},
\newblock Phys. Rev. Lett. {\bf 86} (2001)   4753--6.

\bibitem{Kutasov1989}
D.~Kutasov,
\newblock {\em String theory and the non-abelian {Thirring} model},
\newblock Phys. Lett. B {\bf 227} (1989)   68--72.

\bibitem{BennettGracey1999}
J.F. Bennett and J.A. Gracey,
\newblock {\em Three-loop renormalization of the {SU($N_c$}) non-abelian
  {Thirring} model},
\newblock Nucl. Phys. {\bf B 563} (1999)   390--436.

\bibitem{AliGracey2001}
D.B. Ali and J.A. Gracey,
\newblock {\em Four loop wave function renormalization in the non-abelian
  {Thirring} model},
\newblock Nucl. Phys. {\bf B 605} (2001)   337--64.

\bibitem{BernardLeClair2001}
D.~Bernard and A.~LeClair,
\newblock {\em Strong-weak coupling duality in anisotropic current
  interactions},
\newblock Phys. Lett. B {\bf 512} (2001)   78--84.

\bibitem{Nienhuis1984}
B.~Nienhuis,
\newblock {\em Critical behaviour of two-dimensional spin models and charge
  asymmetry in the coulomb gas},
\newblock J. Stat. Phys. {\bf 34} (1984)   731--61.

\bibitem{FendleySaleurZamolodchikov1993a}
P.~Fendley, H.~Saleur  and A.B. Zamolodchikov,
\newblock {\em Massless flows. {I}. the sine-{Gordon} and {O(n)} models},
\newblock Int. J. Mod. Phys. A {\bf 8} (1993)   5717--50.

\bibitem{FendleySaleurZamolodchikov1993b}
P.~Fendley, H.~Saleur  and A.B. Zamolodchikov,
\newblock {\em Massless flows. {II}. the exact {S}-matrix approach},
\newblock Int. J. Mod. Phys. A {\bf 8} (1993)   5751--78.

\bibitem{Zamolodchikov1986}
A.B. Zamolodchikov,
\newblock {\em {``Irreversibility''} of the flux of the renormalization group
  in a {2D} field theory},
\newblock Pis'ma Zh. Eksp. Theor. Fiz. {\bf 43} (1986)   565--567,
\newblock JETP Lett. {\bf 43} (1986) 730.

\bibitem{BernardLeclair2002}
D.~Bernard and A.~LeClair,
\newblock {\em Renormalization group for network models of quantum hall
  transitions},
\newblock Nucl. Phys. B {\bf 628} (2002)   442--72.

\bibitem{Zinn}
J.~Zinn-Justin,
\newblock {\em Quantum Field Theory and Critical Phenomena},
\newblock Oxford University Press, Oxford, 1989.

\bibitem{Witten1984}
E.~Witten,
\newblock {\em Non-abelian bosonization in two dimensions},
\newblock Communications in Mathematical Physics {\bf 92} (1984)   455--72.

\bibitem{BernardLeClair1990}
D.~Bernard and A.~Leclair,
\newblock {\em The fractional supersymmetric sine-gordon models},
\newblock Phys. Lett. B {\bf 247} (1990)   309--16.

\bibitem{LudwigWieseUnpublishedProgress}
A.W.W. Ludwig and K.J. Wiese,
\newblock unpublished and work in progress.

\bibitem{WieseDavid1995}
K.J. Wiese and F.~David,
\newblock {\em Self-avoiding tethered membranes at the tricritical point},
\newblock Nucl. Phys. {\bf B 450} (1995)   495--557.

\bibitem{WieseDavid1997}
K.J. Wiese and F.~David,
\newblock {\em New renormalization group results for scaling of self-avoiding
  tethered membranes},
\newblock Nucl. Phys. {\bf B 487} (1997)   529--632.

\bibitem{WieseHabil}
K.J. Wiese,
\newblock {\em Polymerized membranes, a review}.
\newblock {\em {\em Volume}~19} of {\em Phase Transitions and Critical
  Phenomena}, Acadamic Press, London, 1999.

\bibitem{LudwigWieseUnpublished}
A.W.W. Ludwig and K.J. Wiese,
\newblock unpublished.

\bibitem{vanRitbergenVermaserenLarin1997}
T.~van Ritbergen, J.A.M. Vermaseren  and S.A. Larin,
\newblock {\em The four-loop beta-function in quantum chromodynamics},
\newblock Phys. Lett. B {\bf 400} (1997)   379--84.

\end{thebibliography}

\end{document}